\documentclass[twocolumn,preprintnumbers,amsmath,amssymb]{revtex4}
\input{epsf}

\newcommand{\be}{\begin{equation}}
\newcommand{\ee}{\end{equation}}
\newcommand{\bea}{\begin{eqnarray}}
\newcommand{\eea}{\end{eqnarray}}
\newcommand{\bef}{\begin{figure}}
\newcommand{\eef}{\end{figure}}

\newcommand{\simge}{\,{}^>_{\sim}\,}
\newcommand{\simle}{\,{}^<_{\sim}\,}

\def\h#1{$^{#1}$H}
\def\he#1{$^{#1}$He}
\def\li#1{$^{#1}$Li}
\def\be#1{$^{#1}$Be}

\def\eps@scaling{0.96}

\def\showone#1{
  \centering
  \leavevmode
  \epsfxsize=\eps@scaling\linewidth
  \epsfbox{#1.eps}
}

\def\epstwo@scaling{0.48}

\def\showtwo#1#2{
  \centering
  \leavevmode
  \epsfxsize=\epstwo@scaling\linewidth
  \epsfbox{#1.eps} \hfil
  \epsfxsize=\epstwo@scaling\linewidth
  \epsfbox{#2.eps}
}

\begin{document}
%@twocolumnfalse\endcsname

\title{The cosmic \li6 and \li7 problems and BBN with long-lived 
charged massive particles}
\author{Karsten Jedamzik} 
\affiliation{Laboratoire de Physique Math\'emathique et Th\'eorique, C.N.R.S.,
Universit\'e de Montpellier II, 34095 Montpellier Cedex 5, France}

\begin{abstract}
Charged massive particles (CHAMPs), when present during the Big Bang 
nucleosynthesis (BBN) era, may significantly alter the synthesis of
light elements when compared to a standard BBN scenario. 
This is due to the formation of bound states with nuclei. This paper
presents a detailed numerical and analytical
analysis of such CHAMP BBN. 
All reactions important for predicting light-element yields are
calculated within the Born approximation. Three priorly neglected
effects are treated in detail:(a) 
photodestruction of bound states due to electromagnetic cascades 
induced by the CHAMP decay, (b) late-time efficient
destruction/production of\h2, \li6, and \li7 due to reactions on
charge $Z=1$ nuclei bound to CHAMPs, and (c) CHAMP exchange 
between nuclei. Each of these effects may induce orders-of-magnitude
changes in the final abundance yields. The study focusses on the impact
of CHAMPs on a possible simultaneous solution of the \li6 and \li7
problems. It is shown that a priorly suggested simultaneous solution 
of the \li6
and \li7 problems for a relic decaying at $\tau_x\approx 1000\,$sec
is only very weakly dependent on the relic being neutral or charged,
unless its hadronic branching ratio is $B_h\ll 10^{-4}$ very small.
By use of a Monte-Carlo analysis it is shown that within CHAMP BBN
the existence of further parameter space for a simultaneous
solution of the \li6 and \li7 problem for long decay times 
$\tau_x\simge 10^6$sec
seems possible but fairly unlikley. 
\end{abstract}

%\pacs{PACS numbers:}

\maketitle

\section{Introduction}

Big Bang nucleosynthesis (BBN) is one of the standard pillars
of modern cosmology. In its simplest version, reduced
to a model with only one parameter, i.e. the contribution
of baryons to the critical density, $\Omega_bh^2\approx 0.0224$~\cite{WMAP}, 
standard BBN predicted and observationally inferred
primordial light element abundances are very close. This holds
particularly true for \h2, and with somewhat less confidence also
for \he4. However, when the $A>4$ elements are considered 
agreement is less convincing. The observationally inferred
\li7/H ratio is about a factor three smaller than that predicted
in SBBN~\cite{li7}. Moreover, \li6 which is known to only be synthesized
at the level \li6/H $\sim 10^{-15}-10^{-14}$ during SBBN
has been recently observed in about a dozen metal-poor halo stars
with abundance \li6/H $\sim 3-5\times 10^{-12}$~\cite{li6,li6old}. 
It is tantalizing
that these observations indicate a plateau-structure, similiar
to that observed in \li7, i.e. \li6 abundance independent of
metallicity of the star, for stars at the lowest metallicities. 
A \li6 plateau, should point to a pregalactic or primordial origin of
this isotope, since the \li6 had already been in place before
stars produced metallicity (and cosmic rays). However, it is cautioned
that fairly uncertain stellar pre-main-sequence (PMS) destruction 
of \li6 could contrive to give an apparent plateau~\cite{richardpriv}.

\li7 (as well as \li6) are observed in the atmospheres of 
metal-poor halo stars. When transported to the hotter interior
of the star, by either convection or turbulence, both isotopes
may be destroyed. It is thus possible that atmospheric
\li7 has been depleted by some factor though
standard stellar models do not forsee this. A number of groups
have recently re-studied this 
possibility~\cite{Richard,li7depletion,Korn}. Postulating
stellar turbulence with a parametrised magnitude, but of unknown origin,
Korn {\it et al}~\cite{Korn} claim that a star-to-star homogeneous
factor 1.95 depletion is possible
and even favorable when observations of the metal-poor
globular cluster NGC6397 are
considered. If true, the remaining factor $\sim 1.5$ could be either
due to systematic errors in the effective stellar temperature calibration
or due to an overestimate of the SBBN predicted \li7 abundance due to
systematic errors in nuclear reaction data. Concerning the second
possibility, a recent remeasurement
of the key \li7 producing reaction (\he3$(\alpha,\gamma)$\be7) seems
to rather indicate a slight underestimate of the synthesized 
\li7~\cite{Confortola}.

\li6 is known to be produced by spallation 
($p + {\rm CNO\to LiBeB}$)
and fusion ($\alpha + \alpha\to$ Li) reactions  
by standard cosmic ray primaries scattering
off nucleons and nuclei in the
intergalactic medium~\cite{CRli6}. Though this process may explain
the observed \li6 at solar metallicity, it is clear, however, 
that it falls short
by a large factor ($\sim 50$) to explain the \li6 observed 
at low metallicity. Similiar holds true for putative cosmic ray
populations due to shocks developed during structure formation~\cite{Inoue}.
In order to produce \li6/H$\sim 5\times 10^{-12}$ an early 
cosmic ray population of $\sim 100$eV/nucleon is required~\cite{Prantzos}.
Most candidate sources fall short of this. The few viable remaining
sources are due to accretion on the central Galactic black 
hole, albeit
with an efficiency a factor $10^4$ larger than that
presently observed, or due to a significant fraction $\sim 0.1$
of all baryons forming supermassive stars (and cosmic rays)~\cite{Prantzos}.
It may also be that our galaxy was host to a radio-loud quasar
some time ago~\cite{Nath}. The energetic problem becomes even exaggerated
when likely \li6 destruction during the stellar PMS~\cite{richardpriv} and 
putative \li6 destruction during the stellar main-sequence
phases are considered, possibly solving the \li7 discrepancy.
Finally, it has also been suggested that the \li6
may result in situ from production by solar flares
within the first billion of years of the star's life~\cite{Tatischeff}. 
Though this seems possible, it is hard to evaluate if a
sufficient fraction of the freshly synthesized \li6 falls back into the
stellar atmosphere rather than being expelled by the solar wind. 
%Nevertheless,
%the observed halo stars showing \li6 do not show any signs of the
%magnetic activity, stellar winds, and/or rapid rotation required to
%support such a scenario.

It is entirely possible that the \li7 and \li6 anomalies are signs
of physics beyond the standard model possibly connected to the quest
for the cosmic dark matter. Even very small non-thermal perturbations
in the early Universe may lead to a significant and observable
\li6 abundance, without overly perturbing other light elements. 
It had thus been suggested that an anomalous high \li6
abundance is due to non-thermal nuclear reactions 
(i.e. \h3 $(\alpha ,n)$\li6 , ...) induced by the late-time $t\simge 10^7$s
electromagnetic~\cite{jeda1,remark3} or hadronic~\cite{DEHS} 
decay of a relic particle, as for example the 
gravitino. 
\li6 in abundance as observed in old stars may also be
synthesized due to residual dark matter annihilations 
during the BBN epoch~\cite{jeda2}. 
In particular, a standard thermal 
freeze-out process of weak scale particle dark matter (such as supersymmetric
neutralinos) is concommitant with the production of 
\li6 in the right amount, given the dark matter mass falls in the
range $20\,\simle m_{\chi}\,\simle 90\,$ GeV, and annihilation is to
a significant fraction hadronic and s-wave. Concerning a solution to the
\li7 problem, early attempts utilising the electromagnetic decay of a relic
and the induced \be7 photodisintegration~\cite{Feng} 
(\li7 is mostly synthesized as \be7,
which later on electron-captures) have not proven viable due to 
unacceptable perturbations in the \h2/H and \he3/\h2 
ratios~\cite{Ellis1}. However, it
has been shown that the hadronic decay of a relic during BBN, and the
induced excessive neutron abundance may prematurely convert \be7 to \li7 which
is then destroyed by proton capture. When $\Omega_{\chi}B_h\sim 1-5\times
10^{-4}$, where $B_h$ is the hadronic branching ratio, 
a factor $2-4$ destruction of \li7 results~\cite{jeda3}. 
For relic decay times 
$\approx 1000\,$s,
it is moreover possible to synthesize all the observed \li6 by non-thermal
nuclear fusion. This has been the first, and so far only, known 
simultaneous solution to the \li6 and \li7 problems. It is noted that
such a decay also leads to a possibly problematic
30\% - 50\% increase in the synthesized \h2/\h1 ratio.

Within the context of minimal supersymmetric extensions of 
the standard model of
particle physics, a simultaneous solution is nicely realised, either by
heavy gravitino decay, or in the case that gravitinos are the 
lightest supersymmetric particles (LSPs) by the supersymmetric partner of the 
tau-lepton (the stau) decaying into gravitinos~\cite{jeda3}. 
In the second scenario, an
added benefit is that for the right parameters to solve the \li6 and \li7 
problems, TeV staus left over from a thermal freeze-out at 
higher temperature,and decaying at $\tau_x\approx 1000\,$s into 
$50-100\,$GeV gravitinos 
produce naturally about the right amount of gravitinos to explain
the dark matter and of a warmness interesting to the formation of large scale
structure formation~\cite{jeda4}. 
Unfortunately, staus of mass $1\,$TeV are too heavy to be discovered
at the LHC.

Recently, it has been realised that the existence of electrically
charged massive particles (CHAMPs) during the BBN epoch may lead to 
modifications of the synthesis of light 
elements~\cite{Pospelov,Kohri,Kaplinghat} beyond those simply due to their
decay. Since for gravitino LSPs,
the next-to-LSP (NLSP) is long-lived and
in about half of the supersymmetric parameter space it is the electrically
charged stau, such effects are important to consider. Other metastable
charged relic particles possibly existing during BBN have been also
proposed~\cite{Fargion}.  
Modifications to BBN
occur due to the formation of electrically
bound states between the negatively charged CHAMPs and the positively
charged nuclei. The realization that (meta)-stable weak-scale
mass charged particles enter into bound states during and after BBN
had already been made in the late 
eighties~\cite{DeRujula,Dimopoulos:1989,Rafelski}, when the possibility of
charged dark matter was analyzed. Nevertheless, the influence of bound
states on BBN had not been much discussed.

In this paper results of the up-to-now most 
detailed calculations of BBN nucleosynthesis
in the presence of decaying negatively charged particles are presented. 
The
analysis attempts to reveal all key processes important for a reliable
prediction of light element yields,
thereby revealing, heretofore neglected effects, which
make orders of magnitude 
changes in the predicted BBN yields for much of the parameter space.
These changes are found mostly for late decaying $\tau_x\simge 10^6$s
CHAMPs. The aim of the paper is to analyze the potential
of bound-state nucleosynthesis to solve the cosmic \li7 and \li6 problems.

The outline of the paper is as follows.
In Section 2 a discussion/analysis of all priorly suggested solutions to
the \li7 problem within bound-state nucleosynthesis is presented,
whereas in Section 3 details of the present calculations are given.
In Section 4 it is shown that BBN continues to very low temperatures
$T\ll 1\,$keV in the presence of bound states. Section 5 shows that
bound states are efficiently photodisintegrated already at high temperature
due to the decay of the relic. Section 6 stresses the importance of
CHAMP transfer reactions at late times. Finally In Section 7
possible further solutions to
the \li6 and \li7 problems for late-decaying CHAMPs $\tau_x\simge 10^6$sec
are discussed. Section 8 draws the conlusions. An appendix gives some detail
on the determination of reaction rates in the Born approximation.

\section{Bound-state BBN and prior suggested solutions to the \li7 problem}

Modifications to BBN
occur due to the formation of electrically
bound states between the negatively charged CHAMPs and the positively
charged nuclei.
Since bound state binding energies may be appreciable (cf. Table 1),
a significant fraction of \be7 may be captured by CHAMPs at temperatures
as high as $T\simle 30\,$keV, whereas the same occurs 
at $T\simle 10\,$keV for \he4. This may be seen in Fig.~\ref{fig1}, which
shows the fractions $f_i^b = n_{(N_iX^-)}/n_{N_i}^{tot}$
of \be7,\li7,\li6, and \he4 locked up within bound states.
On first sight, the most important effect of bound states during BBN
is a reduction of the Coulomb barrier~\cite{Pospelov,Kohri}. 
Nevertheless, since SBBN is essentially
finished at $T\approx 10\,$keV, Coulomb barrier modifications of 
reactions rates involving
\he4 should be hardly important (even though, ad hoc, speculated otherwise
in Ref.~\cite{Kohri}). However, as shown by Pospelov~\cite{Pospelov} 
there is a
non-trivial catalytic effect on reactions involving photons 
in the final state. SBBN reaction rates involving dipole radiation (E1; e.g.
\he3(\he4$ ,\gamma)$\be7) scale as $\lambda_{\gamma}^{-3}$, whereas reaction
rates forbidden at the dipole approximation but allowed at quadrupole (E2; e.g.
\h2(\he4$,\gamma)$\li6) scale as $\lambda_{\gamma}^{-5}$, where 
$\lambda_{\gamma}$ is the wavelength of the emitted photon. This, in both 
cases is around $\sim 130\,$fm. In the presence of a \he4-CHAMP bound state
the reaction may proceed photonless (e.g., \h2(\he4$-X^- ,X^-)$\li6)
and $\lambda_{\gamma}$ is approximately replaced by the Bohr radius 
$a_{\rm ^4He}$ of the \he4-CHAMP bound system. Since 
$a_{\rm ^4He}\approx 4.8\,$fm (cf. Table 1)
very large enhancement factors of $7\times
10^7$ and $3\times 10^5$~\cite{Pospelov,Cyburt} for the S-factors 
of the \h2 + \he4, and
\he3 + \he4 reactions, respectively, 
have been estimated. A recent more detailed three-body
nuclear reaction calculation of the \h2 + \he4 reaction, has reduced this
estimate by a factor $\sim 10$~\cite{Hamaguchi}.
Such large enhancement factors are important as they lead to excessive
\li6 (and \li7) production for any weak scale charged particles which
are sufficiently long-lived $\tau_x\simge 4\times 10^3$s, unless
$\Omega_X\simle 3\times 10^{-6}$. They have thus been utilised 
to place a stringent upper
limit on the reheat temperature in the early Universe $T\simle 10^7$GeV
in the case when the supersymmetric gravitino exists and when it
is the LSP~\cite{Pradler}. Nevertheless, it seems somewhat premature
to set such upper limits, as the BBN with charged long-lived particles
for decay times $\tau_X\simge 10^6$s had priorly not been investigated
(cf. Section 8).

The putative existence of bound states during BBN 
has also led to a flood of claims
of possible solutions to the \li7 and/or \li6 anomalies.
%It is first noted, that the simultaneous solution of the Li problems
%presented in Ref. , and discussed above, is unaffected by bound states 
%since for decay time 
%$\tau_x\approx 10^3$s bound states are essentially absent.
In Ref.~\cite{Kohri} 
it was realized that significant fractions of the \be7 and \li7
isotopes are within bound states during BBN. This has lead
the authors to arbritrarily enhance certain reactions rates involving mass-7 
element destruction processes by large factors, leading to the claim that
the existence of bound states may solve the \li7 overproduction problem.
However, these claims are, up to now, 
unfounded~\cite{remark10} (see also below). 
In Ref.~\cite{Kaplinghat} 
it was noted that during the decay of $X^-$, when 
residing in a bound state with
\he4, the \he4 nucleus could break up. The resultant energetic \h3 and 
\he3 could then fuse on \he4 to produce \li6, in a similar to what had been
proposed in~\cite{DEHS,jeda1}. Though the suggestion
is correct, the authors calculate
the break-up probability to be very small 
(cf. also Ref.~\cite{Bird}), such that the
\li6 synthesis by catalytic \h2 (\he4-$X^-,X^-$)\li6 is by far dominant.
The analysis of Ref.~\cite{Cyburt} (and Ref.~\cite{Cumberbatch})
essentially confirms the simultaneous solutions
to the \li6 and \li7 problems as given in Ref.~\cite{jeda3,jeda4}, 
even when bound state
effects are included. In Ref.~\cite{Jittoh} 
the case of almost degenerate NLSP 
staus $\tilde{\tau}$ 
and LSP neutralinos $\tilde{\chi}$
has been considered. Here mass splittings smaller than
$\delta m = m_{\tilde{\tau}}-m_{\tilde{\chi}}\simle 1\,$GeV
have been assumed. In this region of $\tilde{\tau}$-$\tilde{\chi}$ 
parameter space,
motivated by the well-known $\tilde{\tau}$-$\tilde{\chi}$
coannihilation region for neutralino dark matter, the stau is relatively
long-lived due to final phase space supression of the decay. It is
claimed, that the \li7 overproduction problem may be solved by internal
conversion of staus in bound states with \be7, to neutralinos, e.g.
($\tilde{\tau}$-\be7)$\to \tilde{\chi}+\nu_{\tau}+$\li7 and the subsequent
destruction of \li7 by protons. It is argued that solutions to the \li7
problem may be found for $\delta m\simle 100\,$MeV even for the smallest
abundances of staus. A more detailed analysis of the \be7-bound state fraction
via the Boltzmann equation shows, however, that only a very small fraction
of \be7 are within bound states, thus making modifications of the \li7 
abundance at low stau-density negligible. At larger stau-densities some
effect may result.

\bef
% type 'grep BoundingBox fig1.eps' and modify
\epsfxsize=8.5cm
\epsffile[50 50 410 302]{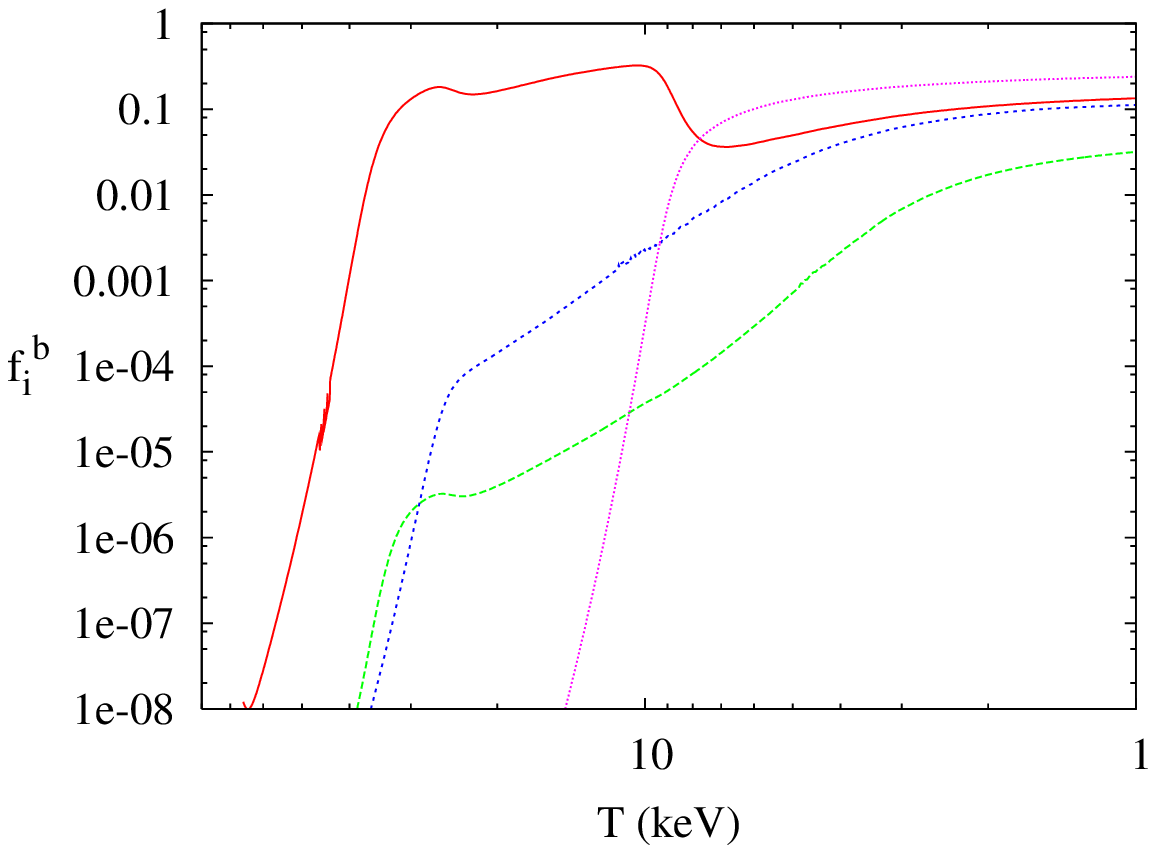}
\caption{Bound state fractions $f_i^b\equiv n_{(N_i X^-)}/n_{N_i}^{tot}$
of nuclei $N_i$ bound to CHAMP $X^-$ as a function of temperature $T$,
for a model with 
$M_X = 100\,$GeV and $\Omega_Xh^2 = 0.1$ (corresponding to a CHAMP-to-baryon
ratio $Y_{X^-}= 4.26\times 10^{-2}/2$). Shown are $f_i^b$ for \be7
solid (red), \li7 long-dashed (green), \li6 short-dashed (blue), and \he4
dotted (purple), respectively. Nuclear destruction of bound states results
in a behaviour of $f_i^b$ different than that expected from simple estimates
by the Saha equation. This is particularly seen in $f_i^b$ for \li7 due
to the \h1(\li7$-X^- ,X^-)$\he4 + \he4 reaction.}
\label{fig1}
\eef

Ref.~\cite{Bird} make the interesting suggestion that the \li7 problem could be
solved by catalytic conversion of \be7 via (\be7-$X^-)(p,\gamma)({^8}{\rm
B}$-$X^-$) and the subsequent 
beta-decay of the ${^8}{\rm B}\to$\be8$+e^++\nu_e$
nucleus. This reaction would mostly occur via a $p$-\be7 resonance in the
${^8}{\rm B}$ nucleus which, in the absence of bound states lies at 769.5keV
relative to the $p$-\be7 continuum. The catalysm in the reaction would then
occur by a shifting of the resonance to $\approx 167\,$keV relative to the
$p$-(\be7-$X^-$) continuum since the (${^8}${\rm B}$-X^-$) bound state binding
energy ($E_{{^8}{\rm B}X^-}\approx 2.0\,$MeV) is larger than that of \be7
($E_{{^7}{\rm Be}X^-}\approx 1.39\,$MeV), making the resonance available at
only slightly supra-thermal energies. Moreover, apart from the decrease in the
resonance energy they also deduce a factor $\sim 10^3$ larger reaction rate
coefficient. Adopting their calculated rates for \be7-$X^-$ bound state
formation and the (\be7-$X^-)(p,\gamma)({^8}{\rm B}$-$X^-$) reaction, I
partially confirm this effect by full numerical analysis.
For example, for $\tau_X = 1.5\times 10^3$s and the number ratio of $X^-$s to
baryons $Y_X\approx 0.2$, I find a reduction of the \li7 abundance by $33\%$,
and a \li6/H ratio of $2\times 10^{-11}$. However, the effect is not as
strong as initially imagined, since by the reciprocity theorem  
the inverse rate is also enhanced. The inverse rate $1/\tau_{\rm inv}$
is thus around $10^3$
times larger at $T\approx 32.5\,$keV than the beta decay rate 
$1/\tau_{\beta}$ of $^8$B (half-time of
$770\,$ ms), converting $^8$B-$X^-$ rapidly back to \be7-$X^- + p$, before
$^8$B can beta-decay. The effect is therefore essentially absent at
early times (i.e. small $\tau_X$). Nevertheless, the inverse rate
quickly drops below the beta decay rate 
(i.e. $\tau_{\beta}/\tau_{\rm inv}\approx 0.1$ at $T\approx 24.1\,$keV).
For the same parameters as above, I still find a $14\%$ reduction of the
final \li7. This drops to $7\%\,$, $2\%\,$ for 
$\tau_X = 10^3$ and $7\times 10^2\,$s, respectively.

It is interesting to know if the solution of the
lithium problems proposed in 
Ref.~\cite{jeda3} is changed when the decaying relic is charged,
such as the stau.
In Fig.~\ref{fig2} the parameter space solving either the \li7 problem,
or both the \li6 and \li7 problems, is shown. The upper panel 
shows results for a charged
relic and the lower panel for a neutral relic. Here observational 
limits as discussed in Ref.~\cite{jeda5} have been applied and the
\li6, \li7 problems are assumed to be reconciled with observational
data for \li6/\li7 $\simge 0.03$ and \li7/\h1 $\simle 2.5\times 10^{-10}$.
The assumed parameters of the model are a hadronic branching ratio
$B_h = 10^{-4}$ and relic mass $M_X = 1\,$TeV.
It is seen that even at $B_h$ as small as $10^{-4}$ the \li7-solving
region is essentially unmodified, whereas some changes are observed in
the \li6 and \li7 solving regions. These latter are mostly due to excessive
\li6 production when the relic is charged, disallowing some of the
larger life times $\tau_X\simge 2\times 10^4$s. 
Bound state effects are nevertheless important when the hadronic
branching ratio is very small. This may be seen in the lowest panel of 
Fig.~\ref{fig2}, where $B_h = 0$ has been assumed. When only bound state
effects are operative, the \h2/\h1-ratio is essentially unmodified. This
is in contrast to the solution of the lithium problems with a hadronic
decay, as seen by the dotted (blue) lines in the upper two panels, beyond
which \h2/\h1 is larger than $4\times 10^{-5}$.
It is intriguing that both processes, hadronic decay and bound state effects,
have the same preferred $\tau_X$ for a simultaneous solution of the lithium
problems.

\begin{footnotesize}
\begin{table}
\newcommand{\lstrut}{{$\strut\atop\strut$}}
\caption{Nucleus, energy of bound state, approximative Bohr radius of bound
state $a_B$~\cite{remark20}, 
and adopted root-mean-square charge radius for nucleus}
\label{T:oh}
\vspace{2mm}
\begin{center}
\begin{tabular}{|c||c|c|c|c|c|}
\hline
 nucleus  & $E_b$ (keV) & $\approx a_B$ (fm) 
& $\langle r^2\rangle^{1/2}_c (fm)$ \\
\hline
\h1 & 24.97 & 28.8 & 0.895 \\
\h2 & 49.5 & 14.4 & 1.3 \\
\h3 & 72.6 & 9.6 & 1.7 \\
\he3 & 269 & 5.2 & 1.951 \\
\he4 & 349.6 & 4.8 & 1.673 \\
\li6 & 842.5 & 2.1 & 2.37 \\
\li7 & 897.6 & 1.9 & 2.50 \\
\be7 & 1385 & 1.5 & 2.50 \\
\hline
\end{tabular}
\end{center}
\end{table}
\end{footnotesize}

\bef
% type 'grep BoundingBox fig1.eps' and modify
\epsfxsize=8.5cm
\epsffile[86 75 319 415]{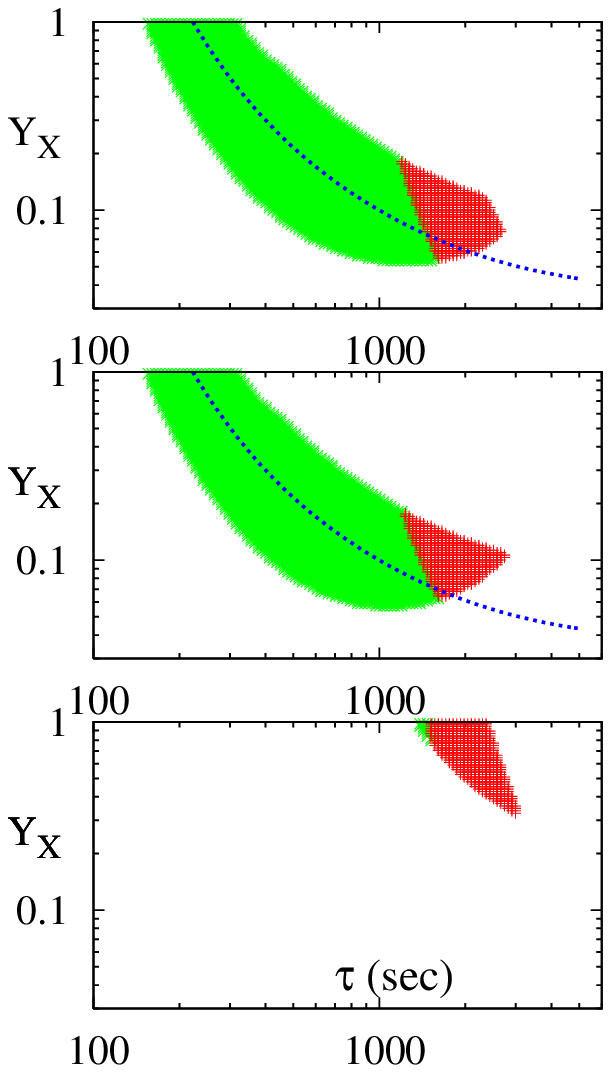}
\caption{Parameter space in the relic particle-to-baryon ratio $Y_X$ and
relic particle decay time $\tau_X$ which may resolve either the \li7
problem (green - light) or both, the \li7 and \li6 problems
(red - dark). The panels show, from top to bottom: (a) a charged relic
with $B_h = 10^{-4}$, (b) a neutral relic with $B_h = 10^{-4}$, and (c)
a charged relic with $B_h = 0$. All three panels assume a mass $M_x = 1\,$TeV
for the relic.
By comparison of the green (lighter) areas it 
is seen that bound-state effects on \li7, as 
suggested in Ref.~\cite{Bird}, 
do not have a very 
big impact for relic hadronic branching ratios 
$B_h\simge 10^{-4}$. The adopted abundance limits are: 
\h2/\h1$<5.3\times 10^{-5}$, \li7/\h1$<2.5\times 10^{-10}$, 
\li6/\li7$<0.66$, and \li6/\li7$>0.03$ to solve the \li6 problem.
Above the dotted lines the \h2/\h1 ratio exceeds a 
value of $4\times 10^{-5}$.}
\label{fig2}
\eef

\section{Detailed bound-state BBN calculations}

The calculations presented here attempt to take proper account
of the influence of singly bound states on the nucleosynthesis for elements
with nucleon number $A\leq 7$. Heavier elements as well as the
formation of molecules, such as $(X^--^4{\rm He}-X^-)$, are not
considered. All effects of electromagnetic and hadronic
cascade nucleosynthesis are included and treated
as presented in Ref.~\cite{jeda5}.  
The fractions of inividual
nuclei $i$ in bound states $f_i^b = n_{(N_iX^-)}/n_{N_i}^{\rm tot}$
are computed by full numerical integration of the
Boltzman equation. This is required since estimates by the Saha equation 
are only very approximative, due to the relatively early freeze-out of the 
CHAMP-nuclei recombination process~\cite{Kohri}. Except of the recombination
rate of $X^-$ on \be7, which is taken from Ref.~\cite{Bird}, 
all other recombination
rates are computed by a numerical integration of the Schroedinger equation. 
This may make difference up to a factor two in $f_i^b$
since the recombination rates as given in Ref.~\cite{Kohri} only apply 
asymptotically at low
temperature $T$. Bound state wave functions and bound-state energies
are also computed by an integration of the Schroedinger equation, assuming
realistic charge radii for the nucleus as measured by experiment. The
reader is referred to Table 1, for some of the bound state 
properties. Finally, it is important, to take into account the nuclear
destruction of bound states. Nuclear rates are very fast at early times,
and for reaction which are sufficiently exothermic, the electric
bound between the final nucleus (nuclei) ought to be 
destroyed~\cite{remark11}.
This often changes $f_i^b$ by orders of magnitude.

\bef
% type 'grep BoundingBox fig1.eps' and modify
\epsfxsize=8.5cm
\epsffile[50 50 410 302]{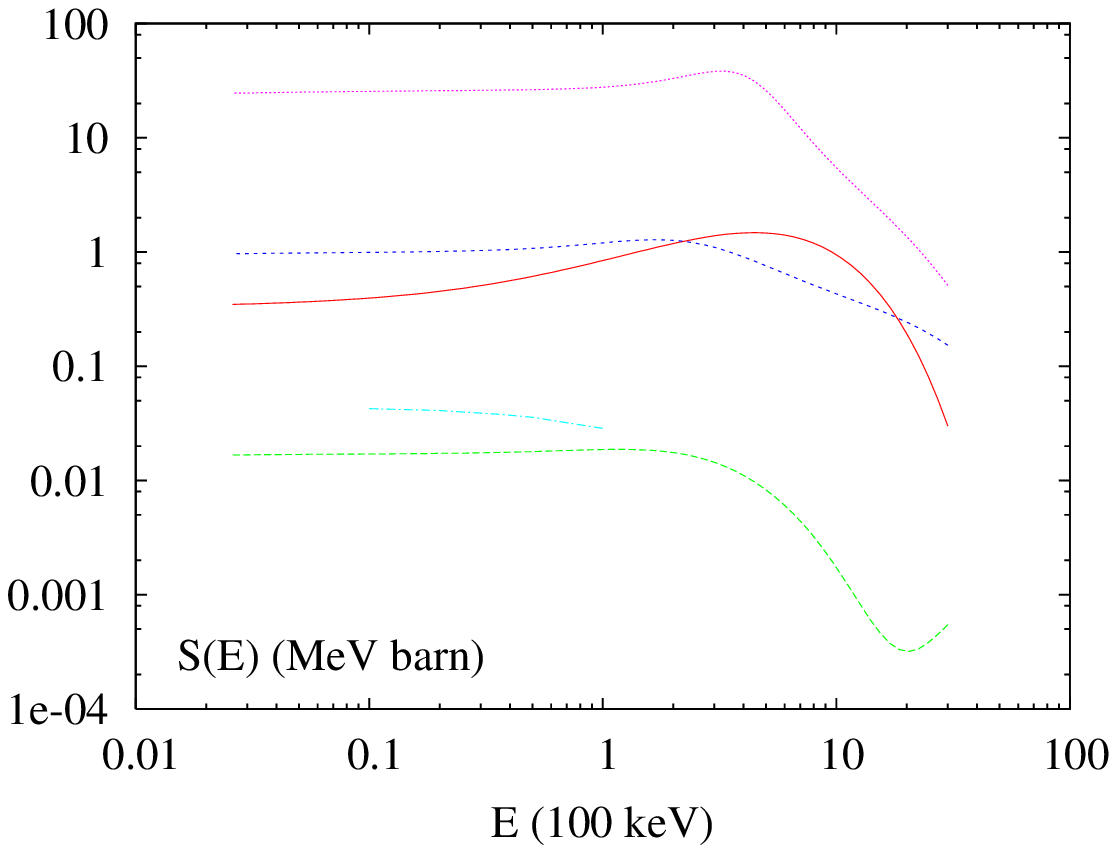}
\caption{Nuclear reaction $S(E)$-factors as function of energy computed
in the present analysis. The most important $S$-factors for nuclear
reactions involving the (\he4-$X^-$)
bound state are shown: \h2(\he4$-X^- ,X^-)$\li6 solid (red), 
\h3(\he4$-X^- ,X^-)$\li7 short-dashed (blue), and
\he3(\he4$-X^- ,X^-)$\be7 dotted (purple), respectively. The dashed-dotted
(light-blue) line shows the result of a recent evaluation~\cite{Hamaguchi} of
\h2(\he4$-X^- ,X^-)$\li6, whereas the long-dashed (green) line shows the
result for the same reaction computed in this paper when the $l=1$ and $l=2$ 
contributions are neglected.}
\label{fig3}
\eef

A proper evaluation of BBN yields with bound states is only possible
when somewhat realistic nuclear reaction rates for nuclei within bound states
are present. With the exception of the reaction \h2(\he4$-X^- ,X^-)$\li6
a more detailed evaluation of such reactions had been absent of the literature
so far. Improving over simple scaling relations~\cite{Pospelov,Cyburt} 
seems important also,
since nuclear reactions including bound states contain three quantities
of similar magnitude, $a_B$ the Bohr ratius, $a_{\rm nucl}$ the nuclear
nuclear radius, and $k_f$ the momentum of the outgoing nucleus.
All three quantities are in the several Fermi range, thus leading potentially
to important cancellation effects. 
More importantly, estimates via simple scaling relations adopt the Born
approximation, which is known to fail at low energies and strong 
perturbations~\cite{remark300}. This is essentially the case for all
reactions of importance to bound-state BBN. The failure of the Born
approximation had been seen, for example, by the
reduction of the \h2(\he4$-X^- ,X^-)$\li6 rate by a factor $\sim 10$, when
a more detailed evaluation~\cite{Hamaguchi} 
is compared to a simple scaling result.

I have identified all key reactions in bound-state BBN. These are shown
in Table 2.
It is completely beyond the scope of the present paper to evaluate
all these reaction rates more properly, i.e. beyond the Born approximation,
a task which is formidable in particular when the important CHAMP-exchange
reactions (cf. Section 6) are also considered.
For the \h2(\he4$-X^- ,X^-)$\li6 process the rate as given by 
Ref.~\cite{Hamaguchi} was adopted.
For other reactions, as a starting point, I have thus
nevertheless, evaluated rates in the
Born approximation. These rates will serve as benchmarks later on.
For details concerning these calculations the reader is referred to
Appendix A.
Results for the in such a
way obtained S-factors are shown in Fig.~\ref{fig3} and Fig.~\ref{fig3a},
respectively.

\bef
% type 'grep BoundingBox fig1.eps' and modify
\epsfxsize=8.5cm
\epsffile[50 50 410 302]{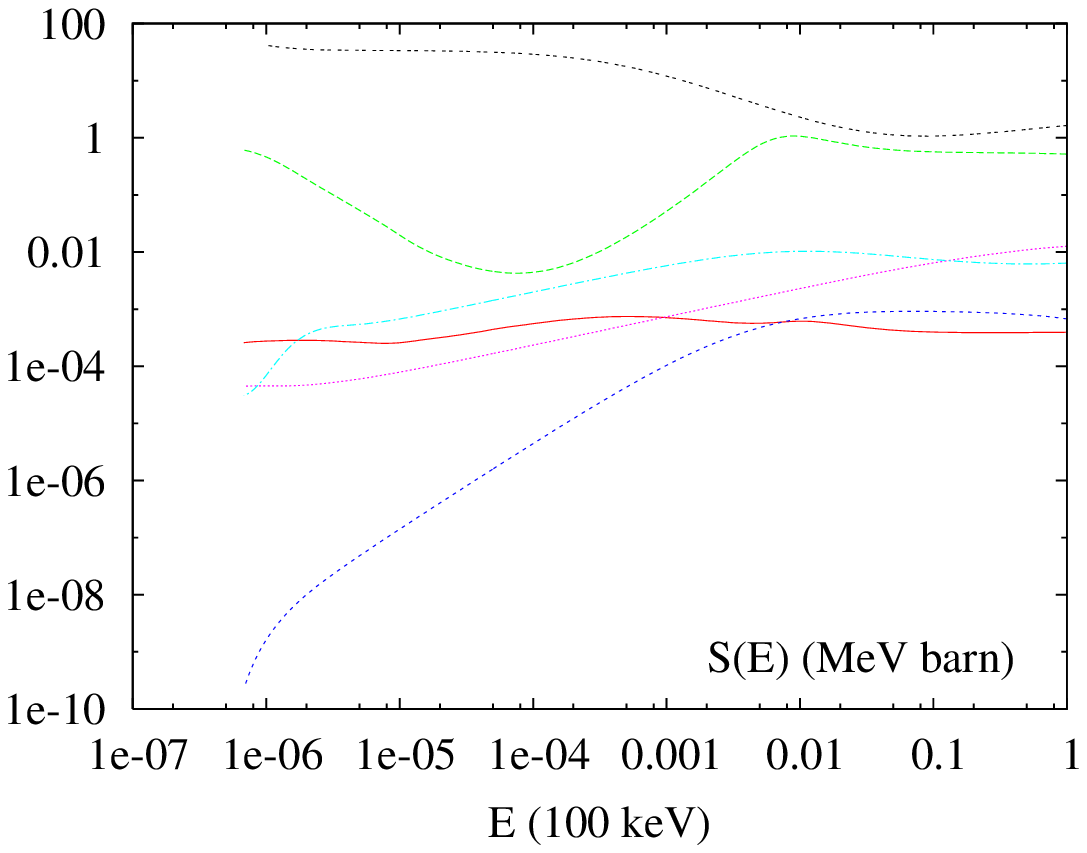}
\caption{
Nuclear reaction $S(E)$-factors as function of energy computed
in the present analysis. The most important $S$-factors for nuclear
reactions involving 
bound states with 
$Z=1$ nuclei are shown: (from top to bottom at the
highest energies) \li6(\h1$-X^- ,X^-)$\he4+\he3 double-dotted (black),
\be7(\h1$-X^- ,X^-){\rm ^8B}$ long-dashed (green),
\he4(\h3$-X^- ,X^-)$\li7 dotted (purple),
\he4(\h2$-X^- ,X^-)$\li6 dash-dotted (light-blue),
\li7(\h1$-X^- ,\gamma )({\rm ^8Be}-X^-)$ short-dashed (blue), and
\li6(\h1$-X^- ,X^-)$\be7 solid (red).
} 
\label{fig3a}
\eef

\begin{footnotesize}
\begin{table}
\newcommand{\lstrut}{{$\strut\atop\strut$}}
\caption{Assumed properties for the calculation of nuclear
reactions with one nuclei in a bound state. The columns show: 
Reaction, $S$-factor for the SBBN reaction
in MeV barn, angular momentum for the $(AB) = C$ final bound nucleus,
and the multipoles
for the initial Coulomb wave which are
included in the calculation.}
\label{T:oh2}
\vspace{2mm}
\begin{center}
\begin{tabular}{|c|c||c|c|c|}
\hline
 No. & $(AX) + B\to C + X$  & $S_{\gamma}$ & $l_C$  & $l^i_{Coul}$ \\
\hline
1 & (\he4-$X^-$) + \h2$\to$ \li6 + $X^-$ & $10^{-8}$ & 0 & 0,1,2 \\
2 & (\he4-$X^-$) + \h3$\to$ \li7 + $X^-$ & $8\times 10^{-5}$ & 1 & 0,1 \\
3 & (\he4-$X^-$) + \he3$\to$ \be7 + $X^-$ & $4\times 10^{-4}$ & 1 & 0,1 \\
4 & (\h1-$X^-$) + \li6$\to$ \be7 + $X^-$ & $10^{-4}$ & 1 & 0,1 \\
5 & (\h1-$X^-$) + \li6$\to$ \he4 + \he3 + $X^-$ & 3 & - & - \\
6 & (\h1-$X^-$) + \li7$\to ({}^8$Be-$X^-$) + $\gamma$ & $10^{-3}$ & 1 & 0,1 \\
7 & (\h1-$X^-$) + \be7$\to {}^8$B + $X^-$ & $3\times 10^{-5}$ & 1 & 0,1 \\
8 & (\h2-$X^-$) + \he4$\to$\li6 + $X^-$ & $10^{-8}$ & 0 & 0,1,2 \\
9 & (\h3-$X^-$) + \he4$\to$\li7 + $X^-$ & $8\times 10^{-5}$ & 1 & 0,1 \\
\hline
\end{tabular}
\end{center}
\end{table}
\end{footnotesize}

\section{Late-time bound-state Big Bang nucleosynthesis}

The reader may have noted that Table 2 also includes reactions with bound
states on elements \h1,\h2,\h3 with only one charge number $Z=1$. In
fact, such reactions are extremely important at low temperatures 
$T\simle 3,2,1\,$keV when one after the other, non-negligible fraction
of \h3, \h2, and \h1 enter into bound states. 
This may be seen in Fig.~\ref{fig5}. It is noted here, that a possible
impact of such reactions has been pointed out before~\cite{Kohri}, albeit in
a very approximative way.
It was not clear, a priori,
if the Coulomb barrier between, for example, \h1 and \li6 is sufficiently
supressed in order to make reactions such as \li6(\h1$-X^- ,X^-)$\he4 + \he3
efficient enough to substantially reduce any priorly synthesized \li6.
This is because, on first sight, 
Coulomb shielding of the
proton could only be partial, due to the fairly extended Bohr radius
$a_B\approx 29\,$fm of the \h1-$X^-$ system. In Fig.~\ref{fig4} 
a $l=0$ spherical
wave without any Coulomb repulsion, i.e. $V_c = 0$,
is compared to the spherical Coulomb wave functions between 
the \li6 and the \h1-$X^-$
bound state with $l=0$ and $l=1$ initial angular momentum, respectively. 
It is seen that essentially no Coulomb supression exists. Rather, the
incoming wave function of the \li6 nuclei is even strongly enhanced
at the center. This is not surprising, as by assumption, the $X^-$ resides
at the center, and due to the significant spread in the wave function of the
proton ($a_B\approx 29\,$fm) the effective proton charge density at the
center is low. The Coulomb potential for the \li6 nucleus is 
$\phi_{{}^6\rm Li} = - 3e^2{\rm exp}(-2r/a_B)(1/r + 1/a_B)$, thus
very attractive at the center and approaching zero at large distances. 
Nuclear reactions between such
bound states and bare nuclei, are therefore not Coulomb supressed.
It is rather conceivable, that Coulomb focussing occurs at low energies,
even enhancing the reaction rates over the $V_C = 0$ case. This may be
observed in the $S$-factor for the (\h1-$X^-$) + \li6$\to$ \he4 + \he3 + $X^-$
reaction as shown in Fig.~\ref{fig3a}. It is noted here, that due to
an anomously low \li7(\h1$-X^- ,X^-){}^8$Be rate found in the Born 
approximation the rate for \li7(\h1$-X^- ,\gamma )({}^8$Be$-X^-$) has
been comuputed and utilised in the calculations.
%One finds thus
%typical cross sections in the barn to $\sim 1000\,$ of barns regime
%at energy $E\approx 1\,$keV. 
%The exception here is the \li7(\h1$-X^- ,X^-){}^8$Be reaction, which is
%found much smaller than that.

\bef
% type 'grep BoundingBox fig1.eps' and modify
\epsfxsize=8.5cm
\epsffile[50 50 410 302]{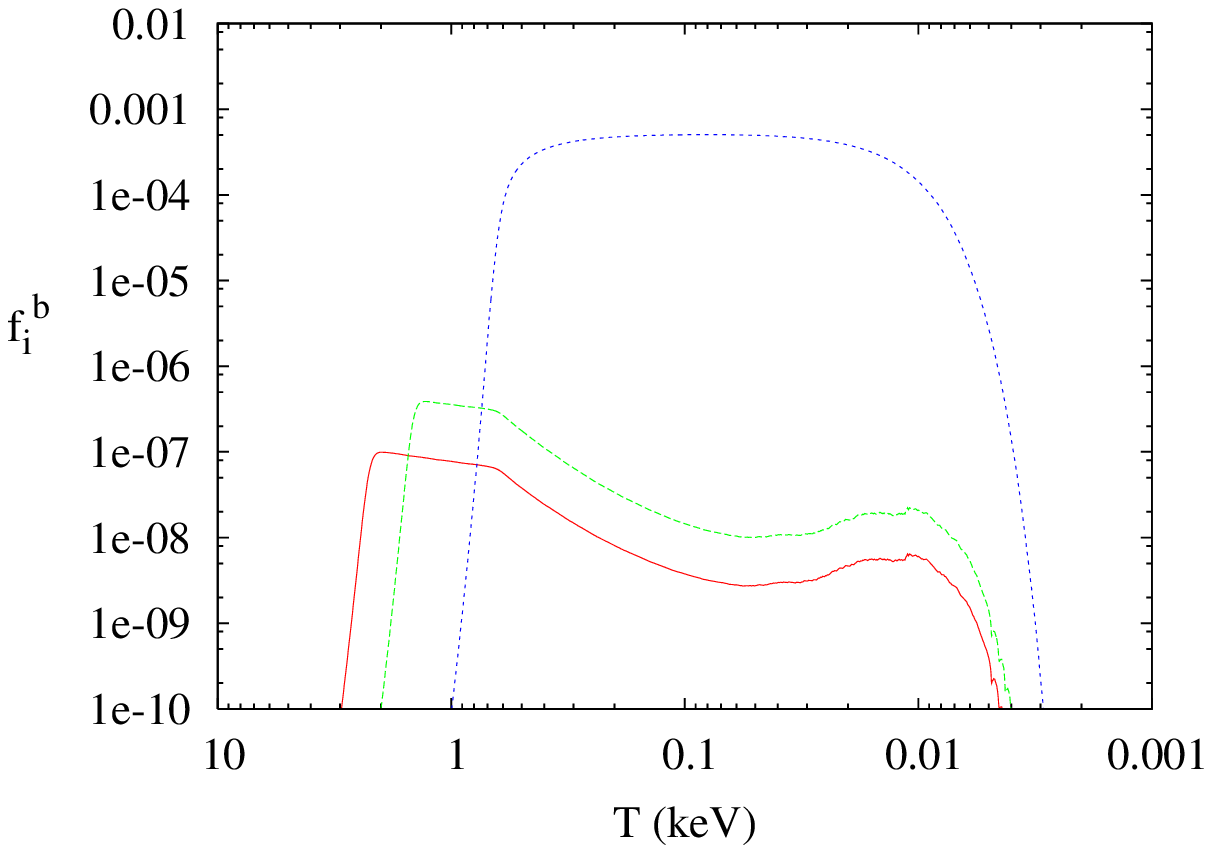}
\caption{Bound state fractions $f_i^b$ for \h3 (solid - red),
\h2 (dashed - green), and \h1 (blue - dotted) as a function of
temperature $T$. Adopted model parameters are as in Fig.~\ref{fig1}
with a $X$ decay time $\tau_X = 10^{10}$s. For illustrative purposes 
photodisintegration of bound states due to $X$-decay (cf. Section 5) and
$X$-exchange reactions (cf. Section 6) have not been taken into account.}
\label{fig5}
\eef

\bef
% type 'grep BoundingBox fig1.eps' and modify
\epsfxsize=8.5cm
\epsffile[50 50 410 302]{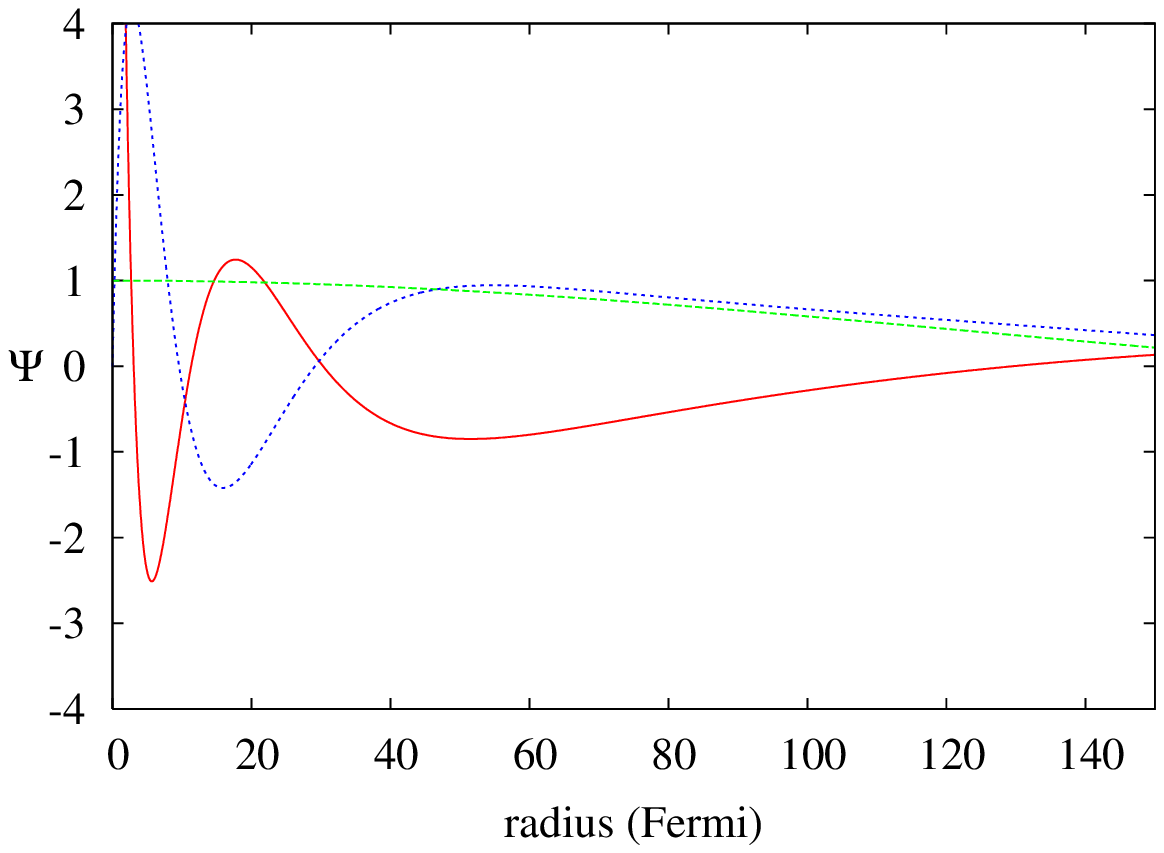}
\caption{Spherical Coulomb wave functions of a \li6 nuclei with energy
$E = 1\,$keV
in the electric field of the \h1-$X^-$ bound state, for s-wave (angular
momentum $l=0$ - solid - red) 
and p-wave (angular momentum $l=1$ - dotted - blue) where $X^-$ is at
radius $r=0$. For comparison the spherical wave function without any
Coulomb barrier, i.e. $V_C = 0$, for s-wave, is also shown (dashed - green). 
It is
seen that no significant Coulomb barrier supression of the wave function
near the origin exists. Rather, both Coulomb wave functions are 
significantly enhanced
at the center, due to the presence of  $X^-$ at $r=0$. 
The oscillatory behaviour may lead to important interference effects.
Both, the $l=0$ and $l=1$ initial states have significant contributions
to the cross section.}
\label{fig4}
\eef

Thus, $Z=1$ bound states at $T\approx 1\,$ keV behave almost as neutrons
(with the exception that they are stable).
Already very small fractions of these bound states
induce therefore a second round of late-time nucleosynthesis,
capable of destroying all the synthesized \li6,\be9, and some of the \li7. 
This may be seen in Fig.~\ref{fig7} where the \li6/H, \li7/H, \be7/H, and \h2/H
ratios are shown for a CHAMP with $\Omega_Xh^2 = 0.01$, 
$m_X = 100\,$ GeV, and decay time
$\tau_X = 10^{10}$s, where $h$ is the dimensionless present-day Hubble
parameter, and $\Omega_X$~\cite{remark12} 
the fractional contribution of CHAMPs to the
present critical density, would they not have decayed. Note, that this
is easily converted to the CHAMP-to-baryon ratio 
$Y_X = (\Omega_X h^2/\Omega_b h^2)(m_p/m_X)$ which is 
$Y_X\approx 4.26\times 10^{-3}$ for the adopted parameters. 
The calculations presented in Fig.~\ref{fig7} 
(as well as Figs.~\ref{fig1} and~\ref{fig5}) are performed 
under the assumption that the $X$ decay is not associated
with any electromagnetic- or hadronic- energy release and in the
absence of $X$-exchange reactions (cf. Section 6). This is done to
isolate the effects of the bound states. At early times, towards the end
of conventional BBN, when a significant fraction of \he4 enters bound states,
the reactions \h2(\he4$-X^- ,X^-)$\li6,
\h3(\he4$-X^- ,X^-)$\li7, and \he3(\he4$-X^- ,X^-)$\be7, 
synthesize significant,
and observationally completely unacceptable abundances of the $A > 4$ isotopes.
However, when bound states of the $Z=1$ elements form at $T\approx 1\,$keV,
essentially all the synthesized \li6 and \be7 may be rapidly destroyed by
the reactions \li6(\h1$-X^- ,X^-)$\he4 + \he3 and 
\be7(\h1$-X^- ,X^-){}^8$B. The situation appears different for \li7,
due to the small estimate for the \li7(\h1$-X^- ,X^-){}^8$Be and
\li7(\h1$-X^- ,\gamma ){}^8$Be-$X^-$ cross sections,
implying that almost all initially synthesized \li7 is left 
intact~\cite{remark13}.
The abundance of \li7/H is found at an observationally friendly
$2.7 \times 10^{-10}$. 
It is noted that \h2 is also destroyed, though to a much smaller degree, 
mostly by the
reactions \h3(\h2$-X^- ,n)$\he4$+X^-$, \he3(\h2$-X^- ,p)$\he4$+X^-$,
and \h2(\h3$-X^- ,n)$\he4+$X^-$, and to a lesser degree by
\h2(\h1$-X^- ,X^-)$\he3. The reader is referred to Table 3 concerning
assumptions about the rate of these, and some other reactions involving
only $A\leq 4$ elements.
Furthermore, when regarding Fig.~\ref{fig7} in more
detail, one also notes late-time 
production of \li6 and \be7 at some level due
to the \he4(\h2$-X^- ,X^-)$\li6 
as well as the \li6(\h1$-X^- ,X^-)$\be7 
reactions.

\begin{footnotesize}
\begin{table}
\newcommand{\lstrut}{{$\strut\atop\strut$}}
\caption{Assumed enhancement factor of a number of nuclear reactions
between $A\leq 4$ nuclei
involving bound states of $Z=1$ nuclei. The Coulomb supression factor
is assumed to be completely absent in these reactions.}
\label{T:3a}
\vspace{2mm}
\begin{center}
\begin{tabular}{|c||c||c|}
\hline
 No. & $(AX) + B\to C + X$  & enhancement \\
\hline
 10  & \h2(\h1$-X^- ,X^-)$\he3 & $1.25\times 10^2$ \\
   & \h1(\h2$-X^- ,X^-)$\he3 &  \\
 11  & \h3(\h1$-X^- ,X^-)$\he4 & 10.7 \\
   & \h1(\h3$-X^- ,X^-)$\he4 &  \\
 12  & \h2(\h3$-X^- ,n)$\he4+$X^-$  & 1 \\
   & \h3(\h2$-X^- ,n)$\he4+$X^-$  &  \\
 13  & \he3(\h2$-X^- ,p)$\he4$+X^-$ & 1 \\
\hline
\end{tabular}
\end{center}
\end{table}
\end{footnotesize}

It is thus premature to conclude, 
that extreme \li6 overproduction, rules out the existence
of CHAMPs with long life times~\cite{Pospelov,Kawasaki}. The model shown
above, at CHAMP densities many (five !)
orders above those already claimed to be
ruled out by \li6 overproduction is observationally viable in all abundances. 
Constraints on the existence of CHAMPs in the
early Universe could therefore, in principle, be much 
milder for long $X^-$ life times than initially predicted. Nevertheless,
they is further important physics entering the calculations discussed
in the next two sections.

\bef
% type 'grep BoundingBox fig1.eps' and modify
\epsfxsize=8.5cm
\epsffile[50 50 410 302]{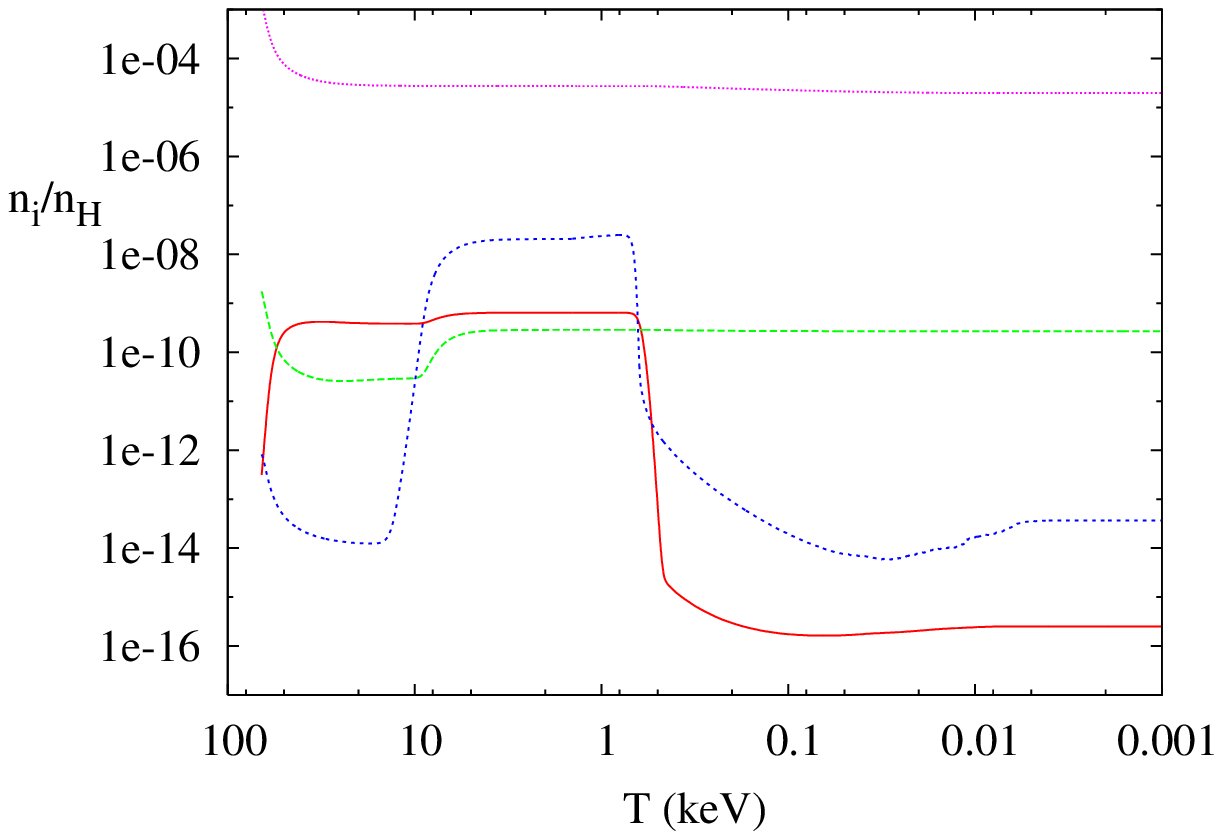}
\caption{Evolution of light-element number ratios \be7/\h1 (solid - red),
\li7/\h1 (long-dashed - green), \li6/\h1 (short-dashed - blue), 
and \h2/\h1 (dotted - purple), for a CHAMP model with $M_X = 100\,$GeV,
$\Omega_Xh^2 = 0.01$, and $\tau_X = 10^{10}$s. 
It is seen that large amounts of \li6 and \be7 synthesized
at $T\approx 10\,$keV will be again destroyed at $T\approx 1\,$keV.
Neither effects due to electromagnetic and hadronic energy
release during CHAMP decay nor charge exchange effects have been
taken into account.}
\label{fig7}
\eef

\section{Photodisintegration of bound states by the decay of the
CHAMPs}

There is another effect, heretofore overlooked, which may significantly 
reduce the in catalytic BBN at $T\approx 10\,$keV
synthesized \li6 (and \li7) abundance. CHAMP decays are typically
accompanied by the injection of electromagnetically interacting partcles,
with total energy comprising often a large fraction of the $X$ rest mass.
It is well-known, that such particles ($e^-,e^+$, and $\gamma$'s) induce
a rapid cascade on the cosmic blackbody photons, 
due to $\gamma\gamma_{\rm BB}$ pair creation and 
inverse Compton scattering $e^{\pm}+\gamma$ processes, until the
energy of any remaining $\gamma$'s is too low to further pair-produce,
i.e. for $E_{\gamma}\simle E_{th}\approx m_e^2/22T
\approx 1.2\,{\rm MeV}(T/10{\rm keV})^{-1}$. It is seen, that
this energy is above the binding energy of \he4-$X^-$ (and \h1-$X^-$) even
at temperatures as high as $T\approx 30\,$keV, making possible the
\he4-$X^-$ and (\h1-$X^-$) bound state photodisintegration before any
significant \li6 synthesis (destruction) has occured. In Fig~\ref{fig8} the
resultant photon spectrum due to the injection of energetic 
electromagnetically interacting particles at cosmic epochs with temperature
$T= 10,1,$ and $0.1\,$keV is shown. The shown spectrum 
$E_{\gamma}\,dn_{\gamma}/d{\rm ln}E_{\gamma}$ is generated by a Monte-Carlo
simulation taking account, not only of $e^{\pm}$ pair production and
inverse Compton scattering, but also $\gamma\gamma$ scattering (important
at high $E_{\gamma}\simle E_{th}$, Bethe-Heitler pair production
$\gamma + {\rm p, {}^4He}\to {\rm p, {}^4He} + e^- + e^+$, Compton scattering
of the produced $e^{\pm}$, as well as the important Thomson (Klein-Nishina)
scattering of $\gamma$'s on thermal electrons. It is based on the calculations
presented in Ref.~\cite{jeda5}, with the Thomson scattering process 
extended to energies
as low as $E_{\gamma}\approx 25\,$keV, to account for \h1-$X^-$ destruction.

Following secondary and tertiary, etc. generations of scattered photons to
obtain the correct photon spectrum for the bound state destructions process
is mandatory. For example, the injection of $1\,$TeV of electromagnetically
interacting energy at $T = 1\,$ keV
is associated with injection of $N_{\gamma}\approx 3.3\times 10^6$ 
primary photons with energy 
$E_{\gamma}\simge 25\,$keV, resulting from the initial
cascade on the blackbody. When further interactions of these $\gamma$'s are
considered the number rises to $N_{\gamma}\approx 1.1\times 10^8$.
In other words, an injected photon takes about 30 interactions before
dropping below the threshold for \h1-$X^-$ photodisintegration. 
This exemplifies the
importance of subsequent $\gamma$ interactions. In Fig.~\ref{fig8} 
one may note a
"pile-up" of photons at low $E_{\gamma}$. This is due to the typical 
fractional loss of $\gamma$'s in the Thomson regime $E_{\gamma}\simle m_e$
being small, such that it takes several Thomson scatterings for
a photon to have dropped below 
$E_{\gamma}\simle E_{\rm ^1H}^b\approx 25\,$keV. 
A similiar pile-up does not exist at 
$E_{\gamma}\simle E_{\rm {}^4He}^b\approx 350\,$keV since during
scatterings of $\gamma$'s with energy $E_{\gamma}\sim m_e$ on electrons
the $\gamma$'s may loose a significant fraction of their energy. We thus
expect the effects of photodisintegration of bound states have a larger
impact on the \h1-$X^-$ bound state fraction than on that of \he4-$X^-$.
This effect is not only due to the above, but also due to the
photodisintegration cross section of \h1-$X^-$, 
$\sigma^{\gamma}_{{\rm {}^1H}-X^-}$ being larger than the
one for \he4-$X^-$. Note that all calculations below, include numerically
evaluated cross sections for the photodisintegration of all $A\leq 7$
nuclei bound states.

\bef
% type 'grep BoundingBox fig1.eps' and modify
\epsfxsize=8.5cm
\epsffile[50 50 410 302]{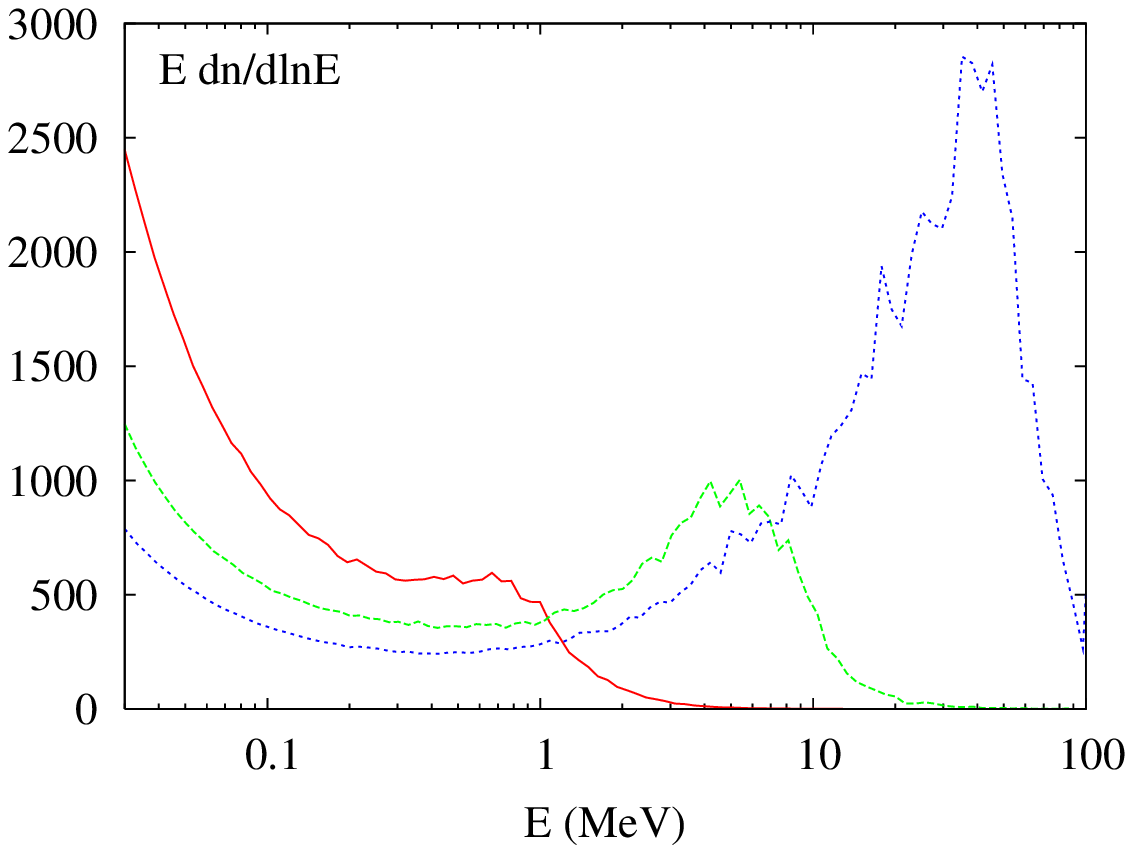}
\caption{Resultant photon spectrum 
$E_{\gamma} {\rm d}n_{\gamma}/{\rm d\,ln}E_{\gamma}$ due to electromagnetic
energy injection at cosmic epochs with temperatures
at $T = 10\,$keV (solid -red), $1\,$keV (dashed - green), and 
$0.1\,$keV (dotted - blue), respectively. The normalisation of the
spectrum is arbritrary. The fraction of photons with energy above the
\he4-$X^-$ photodisintegration threshold $E_4^b\approx 350\,$keV is
$\approx 1.9\%$, $3.7\%$, and $4.7\%$ 
for temperatures $T = 10,1,$ and $0.1\,$keV,
respectively.}
\label{fig8}
\eef

\bef
% type 'grep BoundingBox fig1.eps' and modify
\epsfxsize=8.5cm
\epsffile[50 50 410 302]{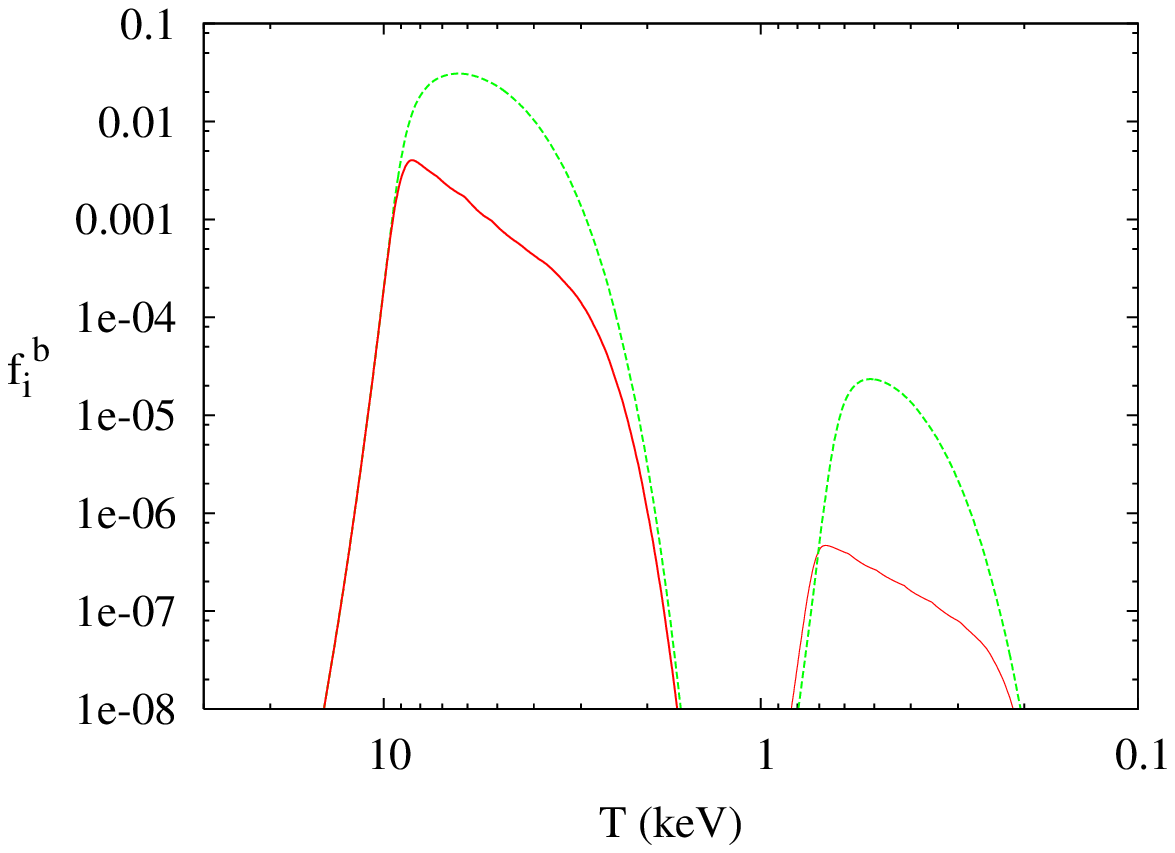}
\caption{\he4 bound state fraction $f_4^b$ for a CHAMP BBN model (A) with
$\Omega_Xh^2 = 0.1$, $\tau_X = 3\times 10^4$s, and $f_{EM} = 1$
(the two curves on the left), and
\h1 bound state fraction $f_1^b$ for a CHAMP BBN model (B) with
$\Omega_Xh^2 = 5\times 10^{-3}$, $\tau_X = 3\times 10^6$s, and $f_{EM} = 1$
(the two curves on the right). Solid (red) curves show $f_i^b$ when
photodisintegration of bound states due to electromagnetic energy release
during the $X$-decay is included, whereas dashed (green) curves show results
when this process is neglected. The resultant \li6 yield in model (A) is
$\sim 10$ times lower than when photodisintegration is excluded. Similarly,
the resultant \li6 yield in model (B) is $\sim 100$ times larger than
without photodisintegration. CHAMP-exchange reactions have not been taken
into account.}
\label{fig9}
\eef

In Fig.~\ref{fig9} the bound state fractions in two scenarios: 
(a) of \he4 for a model with
$\Omega_Xh^2=0.1$ and $\tau_X = 3\times 10^4$s (and electromagnetic decay), 
and (b) of \h1 for 
$\Omega_Xh^2=5\times 10^{-3}$ and $\tau_X = 3\times 10^6$s, are shown in the
same graph. Here the solid lines show $f_{\rm {}^4He}^b$ ($f_{\rm {}^1H}^b$)
when non-thermal bound state photodestruction is included, whereas the
dotted lines show results when it is neglected. It is seen that realistic
bound state fractions are significantly lower. In scenario (a) a \li6/H
ratio $\sim 10$ times lower results, compared to when
photodestruction is neglected,
whereas in scenario (b) the \li6/H ratio is $\sim 100$ times higher.
Here case (b) is affected by a reduced efficiency of
\li6(\h1$-X^- ,X^-)$\he4 + \he3, whereas in case (a) 
the reaction \h2(\he4$-X^- ,X^-)$\li6 is rendered less dominant.
For sufficently high $\Omega_X$, and when thermal photodisintegration is
unimportant, the resultant bound state fraction may
be estimated by a steady state between the recombination rate, i.e.
$\langle\sigma v\rangle_{\rm rec} n_{\rm {}^4He}n_{X^-}$ and the
photodisintegration rate, i.e. 
$\langle\sigma c\rangle_{\rm ph} n_{\rm (^4He-X^-)}n_{\gamma}$.
Here $n_{\rm {}^4He}$, $n_{\rm (^4He-X^-)}$, $n_{X^-}$, and
$n_{\gamma}$ are free
\he4, bound \he4, $X^-$, and nonthermal photon number densities, respectively. 
The nonthermal photon number density $n_{\gamma}$ may be obtained from 
$n_{\gamma}\approx {\rm d}n_{X}/{\rm d}t\,
\tau_{Th}N^{\gamma}_{E_b}$ where
${\rm d}n_{X}/{\rm d}t \approx n_X/\tau_X$ before substantial decay,
$\tau_{Th}$ is the life time of photons against Thomson scattering (i.e. the
typical survival time), and $N^{\gamma}_{E_b}$ 
is the typical number of photons per particle decay
with energy above the photodisintegration
threshold $E_b$ (including secondary generations). 
This, for example at $T = 1\,$keV, 
is approximately $4\times 10^6$ and $1\times 10^8$ for
\he4 and \h1 bound state photodisintegration, respectively, 
per 
$1\,$TeV of electromagnetically interacting energy injected
into the plasma.
It is thus found
\begin{equation}
f_{\rm {}^4He}^b \approx \frac{n_{\rm (^4He-X^-)}}{n_{\rm {}^4He}}\approx 
\frac{\langle\sigma v\rangle_{\rm rec}}{\langle\sigma c\rangle_{\rm ph}}
\frac{\tau_X}{\tau_{Th}}
\frac{1}{N^{\gamma}_{E_b}} 
\end{equation}
It may be noted that this expression, which is valid only for large
$Y_X\simge 10^{-2}$ is independent of the CHAMP-to-baryon ratio,
but dependent on the CHAMP life time.

\section{CHAMP exchange reactions}

\begin{footnotesize}
\begin{table}
\newcommand{\lstrut}{{$\strut\atop\strut$}}
\caption{Rates for CHAMP-exchange reactions computed in the
Born approximation.}
\vspace{2mm}
\begin{center}
\begin{tabular}{|c||c||c|}
\hline
 No. & $(AX) + B\to C + X$  & rate [${\rm cm^3s^{-1}}$] \\
\hline
 14 & (\h1$-X^-$) + \h2$\to$ (\h2$-X^-$) + \h1 & $8.8\times 10^{-15}$  \\
 15 & (\h1$-X^-$) + \h3$\to$ (\h3$-X^-$) + \h1 & $1.4\times 10^{-15}$ \\
 16 & (\h2$-X^-$) + \h3$\to$ (\h3$-X^-$) + \h2 & $1.0\times 10^{-14}$ \\
 17 & (\h1$-X^-$) + \he4$\to$ (\he4$-X^-$) + \h1 & $3.6\times 10^{-17}$ \\
 18 & (\h2$-X^-$) + \he4$\to$ (\he4$-X^-$) + \h2 & $2.9\times 10^{-16}$ \\
 19 & (\h3$-X^-$) + \he4$\to$ (\he4$-X^-$) + \h3 & $8.0\times 10^{-16}$ \\
\hline
\end{tabular}
\end{center}
\label{T:3}
\end{table}
\end{footnotesize}

It has been shown in Section 4 that the existence of only small fractions
$f_p^b\sim 10^{-5}$
of protons in bound states, forming below $T < 1\,$keV, may efficiently
destroy again any priorly synthesized \li6 and \be7. In Section 5 it has
been seen that the efficiency of this destruction may be significanlty reduced
when non-thermal photodestruction of bound states is taken into account.
In this section, a further important process reducing late-time \li6 and \be9
destruction is discussed. CHAMPs in bound states may exothermically
transfer to heavier nuclei
of equal or higher charge. In particular,  (\h1$-X^-$) bound
states could be removed by the
(\h1$-X^-$) + \he4$\to$ (\he4$-X^-$) + \h1 charge exchange process.
Charge exchange reactions turn out to be very important. In Table 4
the most important of these processes are presented.
Rates for these processes were calculated in a very similiar way,
i.e. in the Born approximation, to
those of nuclear reactions involving bound states, as presented in Appendix A.
Here the dipole (quadrupole) operators Eq.~(\ref{Hdipole}) (Eq.~\ref{Hquadro})
replaced by the electromagnetic potential between the bound state and the
heavier nucleus. The same arguments as presented in Section 3 apply
concerning the failure of the Born approximation. In particular, rates
given in Table 4 should be only considered as benchmarks, with the true
rates possibly deviating significantly.

\bef
% type 'grep BoundingBox fig1.eps' and modify
\epsfxsize=8.5cm
\epsffile[50 50 410 302]{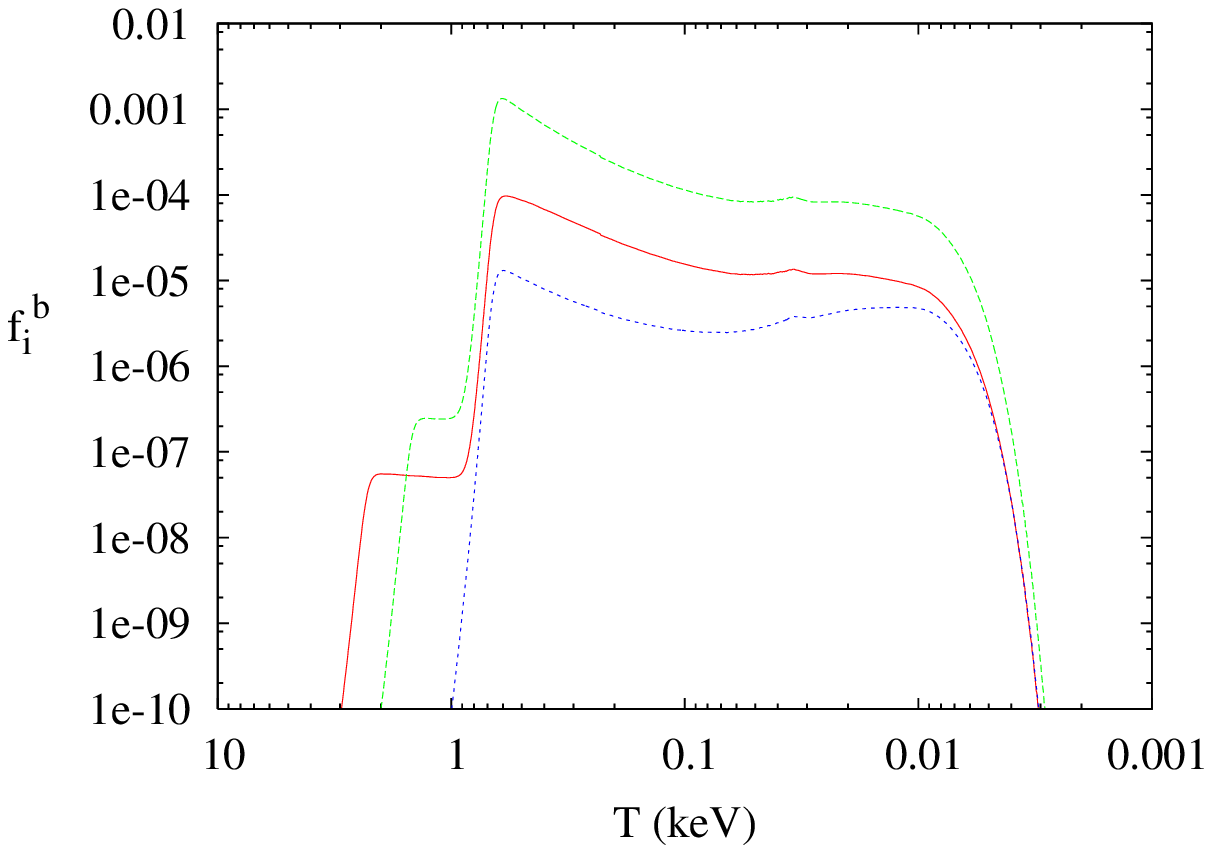}
\caption{As Fig.~\ref{fig5} but with charge exchange reactions included.}
\label{fig5ex}
\eef

\bef
% type 'grep BoundingBox fig1.eps' and modify
\epsfxsize=8.5cm
\epsffile[50 50 410 302]{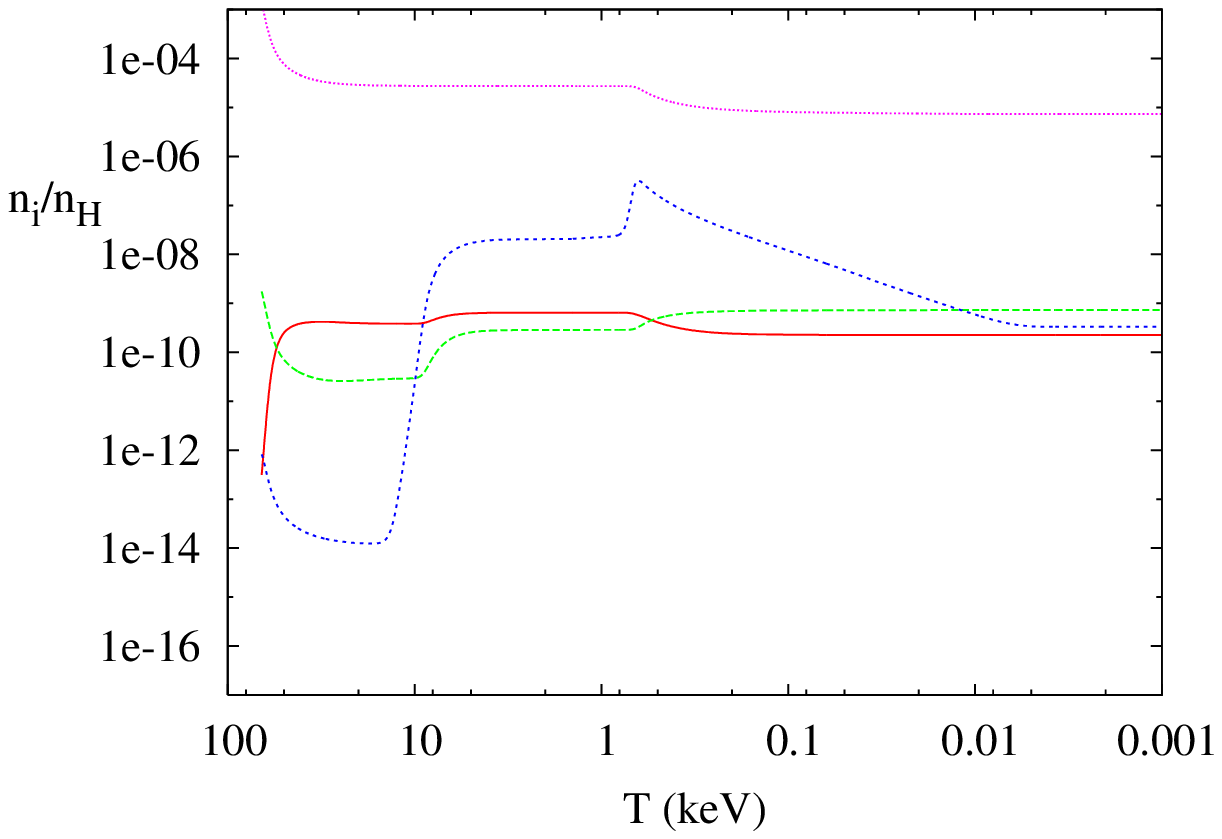}
\caption{As Fig.~\ref{fig7} but with charge exchange reactions
included.}
\label{fig7ex}
\eef

Fig.~\ref{fig5ex} shows bound state fractions for the same model as
that shown in Fig.~\ref{fig5}, but now with CHAMP exchange reactions
included (photodisintegration of bound states is neglected). 
From the comparison of these two figures it is evident that whereas
bound state fractions of \h1 in the absence of exchange reactions
reach levels close to $f_p^b\approx 10^{-3}$, they are two orders of magnitude
below when exchange reactions are present. This is mostly due to
the (\h1$-X^-$) + \he4$\to$ (\he4$-X^-$) + \h1 reaction. 
A for the final BBN yield almost
equally important change is the elevated \h2 (and \h3) bound state fraction
when the reactions in Table 4 are included. Though most \h1 exchange their
CHAMPs with \he4, due to the large \he4 abundance, a large
fraction $\sim 1$ of \h2 enter bound states by capture of CHAMPs
from protons as well. 
The \h2 bound state fraction in Fig.~\ref{fig5ex} (as well as
Fig.~\ref{fig7ex}) is only small $f_D^b\ll 1$, simply because once a
\h2 (and \h3) bound state has formed, its life time against destruction
by reactions shown in Table 3, is very short. In other words, essentially
each \h2 which enters a bound state will be subsequently destroyed, 
leading to the production of \he3 and \he4. This will have important
consequences for bounds on CHAMPs at larger CHAMP density, since either
the lower bound on \h2 or the upper bound on \he3/\h2 may be violated.
Fig.~\ref{fig7ex} shows the abundance evolution corresponding
to Fig.~\ref{fig5ex}, and is the equivalent to
Fig.~\ref{fig7} but now with exchange reactions switched on. 
Several trends are visible: With charge exchange reactions the \h2/H
ratio has fallen below the observational lower limit, i.e.
$7.4\times 10^{-6}$ compared to $2\times 10^{-5}$ in Fig.~\ref{fig7},
the final \li7/H ratio is larger, i.e. $9.5\times 10^{-10}$ compared to
$2.7\times 10^{-10}$, and the \li6/H ratio is much larger, i.e.
$3.3\times 10^{-10}$ compared to $\sim 4\times 10^{-14}$. 
Here \li7 is larger due to reduced \be7(\h1$-X^- ,X^-){}^8$B
and enhanced
\he4(\h3$-X^- ,X^-){}^7$Li efficiencies, and \li6 is larger
due to reduced \li6(\h1$-X^- ,X^-)$\he4 +\he3 and enhanced
\he4(\h2$-X^- ,X^-){}^6$Li reactions.

When doing bound-state BBN computations with reactions on $Z=1$ bound states
as well as CHAMP exchange reactions included often very 
counter-intuitive results are obtained. As only one example, when the
(\h1$-X^-$) + \h2 rate is increased
the \li6 (and \be7) abundance may be reduced
drastically. This is not what is expected since
a lower (\h1$-X^-$) and higher (\h2$-X^-$) fraction 
ought to lead to a higher \li6 
abundance via enhanced \he4(\h2$-X^- ,X^-){}^6$Li and reduced
\li6(\h1$-X^- ,X^-)$\he4 +\he3. Nevertheless, this is not what happens,
due to a higher (\h2$-X^-$) fraction more \h2 is destroyed initially,
rendering the \he4(\h2$-X^- ,X^-){}^6$Li less effective at late times.
Since the final abundance yield is given by the balance of the still fast
processes of
\he4(\h2$-X^- ,X^-){}^6$Li production and \li6(\h1$-X^- ,X^-)$\he4 +\he3
destruction at late times less \li6 results. Due to a lower
\li6(\h1$-X^- ,X^-)$\be7 efficiency less \be7 results. 
Late-time bound-state BBN is
very non-linear requiring full numerical integration up to late
times to obtain reliable
predictions.

\section{Solutions to the \li6 and \li7 problems due to
bound-state BBN for long-lived $\tau_x\simge 10^6$sec CHAMPs ?}

\bef
% type 'grep BoundingBox fig1.eps' and modify
\epsfxsize=8.5cm
\epsffile[50 50 410 302]{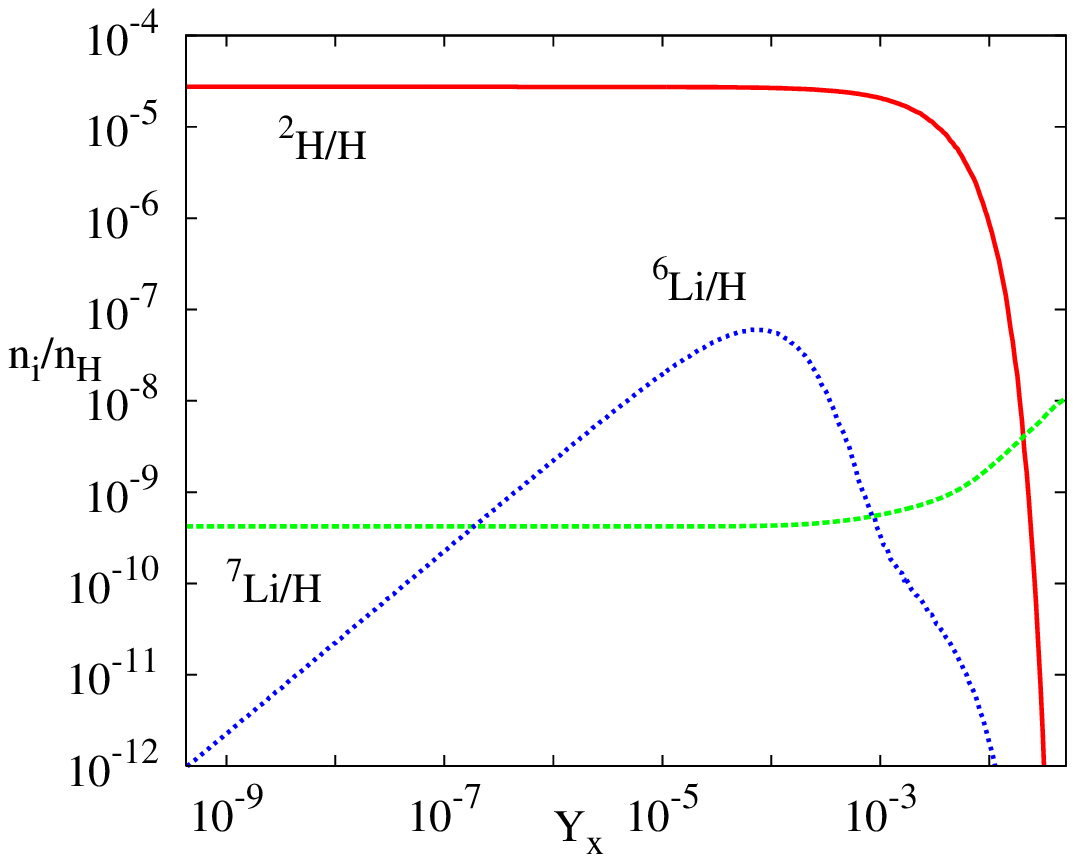}
\caption{Abundance yields of \h2/H solid (red) \li7/H dashed (green),
and \li6/H dotted (blue) as a function of CHAMP-to-baryon ratio $Y_x$
for a model with $\tau_x = 10^{12}$sec and excluding electromagnetic-
and hadronic- energy injection.}
\label{fig21}
\eef

\begin{footnotesize}
\begin{table}
\newcommand{\lstrut}{{$\strut\atop\strut$}}
\caption{Three realizations of models which fulfill constraints
on light-element abundances and reconcile predicted with
observed \li7/\h1 and \li6/\h1 ratios . 
Shown are the CHAMP-to-baryon ratio as
well as a list of reaction numbers and the respective factors by which these
reactions rates have been multiplied with respect to the (unreliable)
estimates in the
Born approximation. All models have $\tau_x = 10^{12}$s and electromagnetic
or hadronic energy injection has not been taken into account, corresponding
to an invisible or almost mass-degenerate decay. Abundance yields in these models
are shown in Table~\ref{T:5}.}
\label{T:4}
\vspace{2mm}
\begin{center}
\begin{tabular}{|c|c|c|}
\hline
 Model & $Y_x$ & Reactions modified \\
\hline
A & $4.3\times 10^{-4}$ & \#4: 0.1 \#7: 2. \#14: 0.1 \#17: 0.3 \\
B & $4.3\times 10^{-4}$ & \#4: 0.3 \#9: 0.3 \#17: 0.1 \#18: 30.  \\
C & $4.3\times 10^{-6}$ & \#5: 3. \#7: 30. \#14: 0.03 \#17: 0.03  \\
\hline

\hline
\end{tabular}
\end{center}
\end{table}
\end{footnotesize}

\begin{footnotesize}
\begin{table}
\newcommand{\lstrut}{{$\strut\atop\strut$}}
\caption{The corresponding abundance yields resulting in the models
shown in Table~\ref{T:4}.}
\label{T:5}
\vspace{2mm}
\begin{center}
\begin{tabular}{|c|c|c|c|}
\hline
 Model & \h2/H & \li7/H & \li6/H \\
\hline
A & $2.6\times 10^{-5}$ & $2.8\times 10^{-10}$ & $9.3\times 10^{-12}$ \\
B & $2.4\times 10^{-5}$ & $2.3\times 10^{-10}$ & $3.7\times 10^{-11}$ \\
C & $2.7\times 10^{-5}$ & $1.5\times 10^{-10}$ & $3.2\times 10^{-11}$ \\
\hline

\hline
\end{tabular}
\end{center}
\end{table}
\end{footnotesize}

\bef
% type 'grep BoundingBox fig1.eps' and modify
\epsfxsize=8.5cm
\epsffile[50 50 410 302]{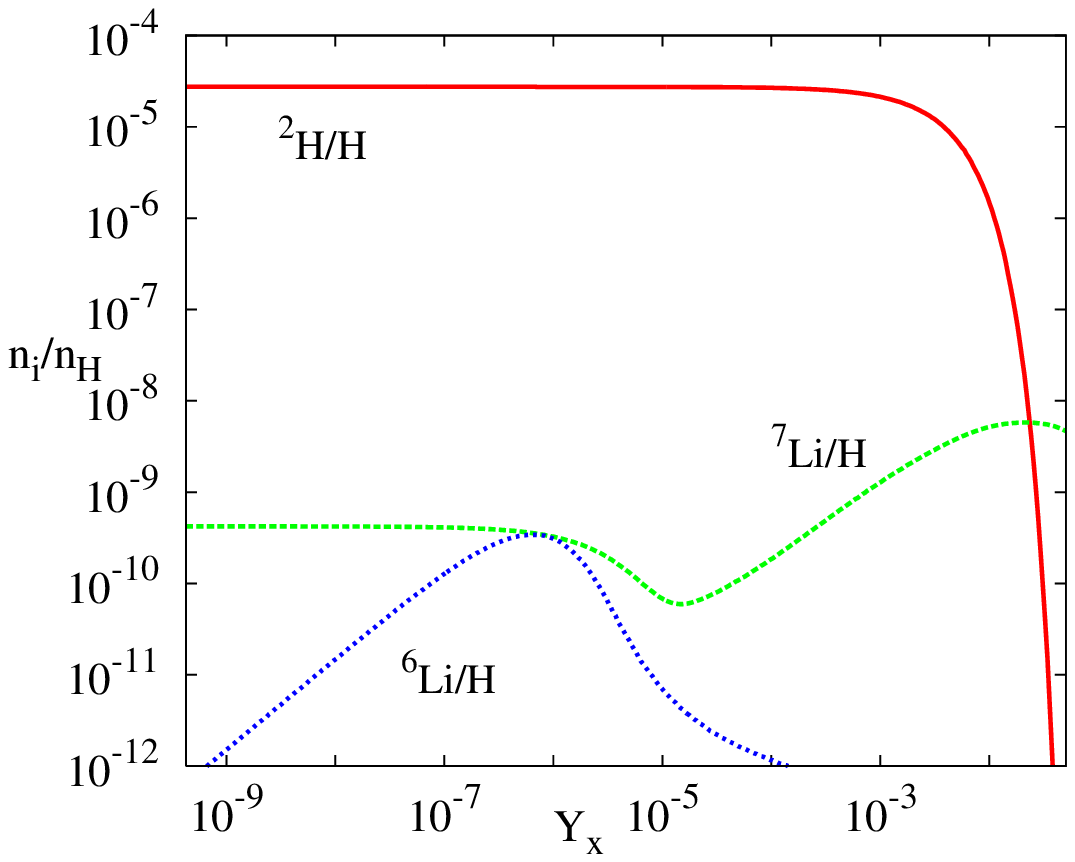}
\caption{As Fig.~\ref{fig101} but for reaction rates as 
in Model C shown in Table~\ref{T:5}.}
\label{fig101}
\eef

In Section 2 priorly proposed solutions to the 
\li7 overabundance and \li6 underabundance
resulting within BBN in the presence of (relatively) short-lived CHAMPs 
have been discussed. Notwithstanding possible
astrophysical explanations of these deviations between theory and observation,
it has been shown that both problems may be solved at once in the presence
of a decaying particle with decay time $\tau_x\approx 1000\,$s. This is
possible in either case, a charged relic or a neutral relic.
In subsequent sections it has been
seen that late-time nucleosynthesis in the presence of charged weak-scale
mass particles may lead to orders-of-magnitude 
modifications of the \li6, \li7, (and \h2) abundances. It would be
interesting to know if CHAMPs with long life times $\tau_x\simge 10^6$s
may reconcile the \li6 and \li7 discrepancies.

In Fig.~\ref{fig21} abundance yields for $\tau_x = 10^{12}$s and
varying $Y_x$ are shown. Here reaction rates in the Born approximation
were adopted 
and electromagnetic- and hadronic- energy injection due
to the $X$ decay was neglected, corresponding to, for example, an invisible
decay or a decay to a neutral daughter particle almost degenerate 
in mass with the CHAMP. The model also approximates well
the case of no decay, i.e. a stable CHAMP.
At low CHAMP-to-baryon ratio $Y_x$ only \li6 is
modified. Here most of the \li6 is synthesized not at early times
due to \h2(\he4$-X^- ,X^-){}^6$Li 
but rather at late times due to \he4(\h2$-X^- ,X^-){}^6$Li. A small CHAMP
density may therefore easily account for \li6 in Pop II stars.
When $Y_x$ increases to $10^{-8}$ too much \li6 is synthesized.
For larger $Y_x\simge 10^{-3}$ \li6 destruction due to
a high (\h1-$X^-$) fraction reduces \li6 again to observationally friendly
levels. However, such models are then ruled out by \li7 overproduction
and \h2 underproduction, due to high (\h2-$X^-$) and (\h3-$X^-$) fractions,
with \li7 produced by \h3(\he4$-X^- ,X^-){}^7$Li and \h2 destroyed by reactions
given in Table 3.
When the decay is electromagnetic or hadronic, with a large
fraction $f_{EM}\sim 1$ of rest mass of $X$ converted
to electromagnetically interacting particles such high $Y_x$ should in any
case be ruled out due to elevated \he3/\h2-ratios (cf. Ref.~\cite{jeda5}).

Nevertheless, significant uncertainties exist due to the uncertainties
in the bound-state nuclear reactions and charge exchange reactions.
In Tables~\ref{T:4} and~\ref{T:5} three (somewhat randomly chosen) models
which do solve the \li6 and \li7 problems are shown. Here a number of
reaction rates were scaled up (or down) from the Born approximation in
order to arrive at an observationally satisfying result. It is seen that
even at low $Y_x$ such models may exist, depending on the exact magnitude
of rates for a variety of reactions. 
It is also seen that, when going to lower $Y_x$,
rates have to deviate more drastically from the Born approximation in
order to solve the \li6 and \li7 problems. The corresponding abundance
yields for Model C, where at low $Y_x$ observationally satisfying results
are obtained, are shown in Fig.~\ref{fig101}. The
figure clearly
indicates that parameter space for a reduction of \li7 and production
of some \li6 exists.

\begin{footnotesize}
\begin{table}
\newcommand{\lstrut}{{$\strut\atop\strut$}}
\caption{Adopted values for $f_i^{cut}$ for the different reactions
varied in the Monte-Carlo analysis (see text for details).}
\label{T:17}
\vspace{2mm}
\begin{center}
\begin{tabular}{|c|c||c|c|}
\hline
 Reac. $i$ & $f_i^{cut}$  & Reac. $i$ & $f_i^{cut}$ \\
\hline
 1  & 3 & 11 & 30 \\
 2  & 30 & 12 & 10 \\
 3  & 30 & 13 & 10 \\
 4  & 30 & 14 & 100 \\
 5  & 30 & 15 & 100 \\
 6  & 30 & 16 & 100 \\
 7  & 30 & 17 & 100 \\
 8  & 30 & 18 & 100 \\
 9  & 30 & 19 & 100 \\
 10  & 30 & & \\
\hline
\end{tabular}
\end{center}
\end{table}
\end{footnotesize}

\bef
% type 'grep BoundingBox fig1.eps' and modify
\epsfxsize=8.5cm
\epsffile[50 50 410 302]{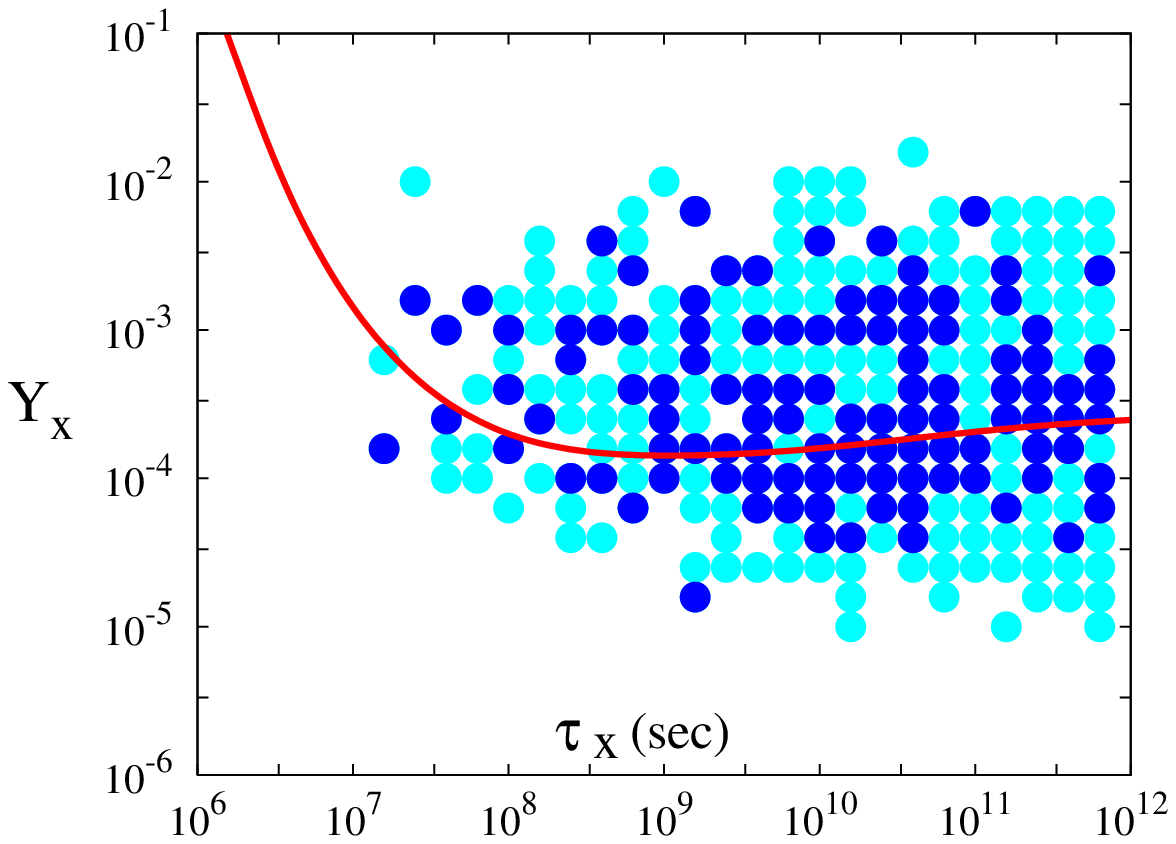}
\caption{Probability in the CHAMP-to-baryon $Y_x$ - CHAMP life time
$\tau_x$ parameter space at large $\tau_x$
that simultaneous solutions to the \li7 and
\li6 (dark-blue) or only \li7 (light-blue) problems exist. The points
indicate $1-5\%$ probability, whereas white areas had less than $1\%$
of all randomly chosen reaction rates in the Monte-Carlo analysis
result in \li6 and \li7 (or \li7 only) solutions. No electromagnetic-
or hadronic- energy release has been taken into account. Areas above the
red line would be ruled out due to electromagnetic cascade nucleosythesis
under the assumption that $f_{EM}=3\times 10^{-2}$ of the rest mass of the
CHAMP is converted to electromagntically interacting energy. 
See text for further detail.}
\label{fig31}
\eef

In the absence of reliable estimates for reaction rates it is difficult
to assess quantitatively
if significant parameter space for 
simultaneous solutions for the \li6 and \li7 discrepancies
for late decaying $\tau_X\simge 10^6$sec CHAMPs exist. 
In particular all
nuclear reactions shown in Table~\ref{T:oh2} and Table~\ref{T:3a}, as well
as the charge exchange reactions shown in Table~\ref{T:3}, i.e. a total number
of nineteen reactions.
Though all rates
have been determined numerically in the Born approximation in this paper,
as the Born approximation is likely to fail badly,
results become uncertain. In order to still arrive at a reliable
result one is thus
forced to perform a Monte-Carlo analysis, varying all ill-determined 
reaction rates within conservative ranges. This has been done in the
present paper. In particular, the Born approximation values of the
rates shown in Figs.~\ref{fig3} and ~\ref{fig3a}, 
as well as given in Table~\ref{T:3a} and~\ref{T:3},
have been taken as benchmarks. For each reaction a random generator
determined a factor $f_i$ with which the benchmark rate was multiplied.
These factors where generated with a probability distribution flat
in logarithmic space, and between values 
$1/f_i^{cut}\leq f_i \leq f_i^{cut}$. For the reaction-rate dependent
conservatively chosen $f_i^{cut}$ the reader is referred to 
Table~\ref{T:17}. For each point in parameter space, i.e. for 
$Y_x$ and $\tau_x$, this procedure was repeated a $1000$ times
in order to arrive with one thousand different randomly chosen sets
for the 19 ill-determined reaction rates. For each realization of
reaction rates an indpendent BBN calculation was then performed
and compared to the observational constraints.

The results of this Monte-Carlo analysis are shown in Fig.~\ref{fig31}.
Here dark (dark-blue) area indicates the probability that between
1\% - 5\% (i.e. 10-50) of all independent 1000 BBN calculations with randomly 
varied rates respect the abundance limits on other light elements 
(as given in Ref.~\cite{jeda5}) while fulfilling \li7/\h1$<2.5\times 10^{-10}$
and $0.66 >$\li6/\li7$> 0.03$. Similarly, light (light-blue) areas
indicate the same, but with now only the \li7 discrepancy solved
(i.e. \li6/\li7$< 0.03$) is acceptable). It is noted that in none of the
parameter space a probability $> 5\%$ for \li6 + \li7 (or only \li7) 
solving areas is found, indicating that the reaction rate combinations
which may yield such solutions are rather rare. The liklihood for such
scenarios is even further diminished when electromagnetic- and/or hadronic-
energy injection due to the decay of the particle is considered. In fact,
when $f_{EM} \sim 1$ all of the parameter space shown in 
Fig.~\ref{fig31} capable of solving
the \li6+\li7 problems simultaneously (though at a $< 5\%$ liklihood),
would be completely eliminated. Only when $f_{EM}$ is rather small, some
area remains. This is shown by the (red) line for $f_{EM} = 3\times 10^{-2}$
corresponding, for example, to the decay of a stau $\tilde{\tau}$ to
a tau and gravitino, with the gravitino only $10\%$ lighter than the stau.
The area above the line is ruled out by overproduction of the \he3/\h2 ratio
due to \he4 photodisintegration. 
On the other hand, not
shown in Fig.~\ref{fig31} are areas where only the \li6 abundance as observed
in Pop II stars may be produced. These exist plentiful, and at high probability,
in particular at lower $Y_x\simle 10^{-5}$. It thus seems unlikley that
CHAMPs with $\tau_x\simge 10^6$sec may resolve the \li7 problem, though they
could possibly constitute the source for the observed \li6 at low metallicity.

\section{Conclusions}

In summary, I have presented results of a very
detailed study of BBN in the presence of negatively 
charged massive particles (CHAMPs). Such particles have been shown
to form bound states with nuclei towards the end of a conventional BBN
epoch
~\cite{DeRujula,Dimopoulos:1989,Rafelski,Pospelov,Kohri,Kaplinghat,Fargion} 
and may alter BBN yields due to the catalysm of nuclear 
reactions~\cite{Pospelov}. 
The present analysis attempts to take into
account of all relevant effects for making relatively precise predictions
of catalytic light-element nucleosynthesis for nuclei with $A\leq 7$,
but excluding the formation of molecules. 
It includes numerical
evaluations in the Born approximation of all 
key nuclear cross sections, where one of the nuclei
is in a bound state. Bound-state recombination and photodisintegration cross
sections are also determined numerically. Furthermore, three very
important and priorly not treated effects for the CHAMP BBN at late
times $\tau\simge 10^5$sec are included: (a) rapid
nuclear reactions including charge $Z=1$ nuclei in bound states, (b)
the photodisintegration of bound states due to $\gamma$- and $x$- rays
generated during the decay of the CHAMPs, and (c) CHAMP-exchange reactions
from a bound state within a lighter nucleus to a bound state within a heavier
nucleus. Light element abundances and bound
state fractions are computed without approximations. The effects of 
hadronic and electromagnetic cascades due to CHAMP disintegration on
light element abundances are properly taken into
account.

The present detailed study reveals that bound-state BBN proceeds
very differently than initially forecasted~\cite{Pospelov,Kohri}. 
At low temperatures $T\simle 1\,$keV, a large number $\sim 20$
of Coulomb-barrier unsupressed nuclear reactions and charge exchange
reactions become operative and are capable, in most of the parameter
space, to change \li6, \li7, and \h2 abundances by orders of magnitude.
Unfortunately, reaction rates for these processes are not well approximated
by the Born approximation, such that for CHAMP life times 
$\tau_x\simge 10^5$sec one has to resort to a Monte-Carlo analysis. 

The purpose of this study is to investigate the
potential of CHAMP BBN to resolve the current \li6 and \li7 discrepancies
between standard BBN and observations. It is shown, that
a priorly proposed simultaneous solution of the \li6 and \li7 problems
with a relic particle decaying at $\tau_x\approx 1000\,$sec~\cite{jeda3},
is not very dependent on the decaying relic being 
charged~\cite{Bird} or not,
unless its hadronic branching ratio is well below $B_h\simle 10^{-4}$. 
A solution with $B_h\ll 10^{-4}$ has, however, the advandtage to not
change much the \h2/\h1 ratio from its respective standard BBN value.
Since \li6 and \li7
may be rapidly destroyed at late times one generically expects further
simultaneous solutions of the \li6 and \li7 problems for
$\tau_x\simge 10^6$sec. Nevertheless, even
given the current
reaction rate uncertainties, a Monte-Carlo shows that only a
very small fraction $\simle 5\%$ of reaction rate combinations may
lead to such solutions. Since such possible solutions occur at relatively
high CHAMP-to-baryon ratio $3\times 10^{-5}\simle Y_x\simle 10^{-2}$
they are further constrained by the effects of electromagnetic energy
injection and possible \he3/\h2 overproduction, 
requiring the decay to be invisible, or mother and daughter
particle to be somewhat degenerate in mass $\sim 10\%$. On the other hand,
CHAMPs may well be the source of the observed
\li6 at low metallicity.

\vskip 0.1in
I acknowledge helpful discussions with 
and M.~Asplund, S.~Bailly, O.~Kartavtsev, K.~Kohri, 
A.~Korn, G.~Moultaka, M.~Pospelov, J.~Rafelski,
G.~Starkman, V.~Tatischeff, and T.~Yanagida. 
\vskip 0.1in

\appendix

\section{Thermonuclear reactions in the presence of bound states in
the Born approximation}

Consider the three-body system of nuclei $A$, $B$, and CHAMP $X^-$.
Since for weak scale mass CHAMPs and light nuclei
$M_{X}\gg M_A, M_B$, it
is an excellent approximation to assume $X^-$ to be at rest at the
origin, effectively acting as an external potential whichs absorbs momentum
but not energy. The Hamiltonian of the system is then given by
\begin{eqnarray}
H = \frac{1}{2}M_A\dot{\bf r}_A^2 + \frac{1}{2}M_B\dot{\bf r}_B^2 + 
V_C(|{\bf r}_A-{\bf r}_B|) + \nonumber \\
V_{\rm NUC}({\bf r}_A,{\bf r}_B)- \frac{Z_Ae^2}{r_A} - \frac{Z_Be^2}{r_B}\, ,
\label{Htot}
\end{eqnarray}
where ${\bf r}_A$, ${\bf r}_B$ represent the position vectors of nuclei
$A$ and $B$,
$r_A$, $r_B$ their magnitudes, and $Z_Ae$, $Z_Be$ their respective charges.
In Eq.~(\ref{Htot}) the first two terms represent kinetic energies, the
second and third term, Coulomb and nuclear potentials between $A$ and $B$,
and the last terms, the Coulomb potentials between the 
(assumed singly charged) CHAMP $X^-$ and the
nuclei. This Hamiltonian will be split into a dominant contribution $H_0$
and a perturbative contribution $H_1\ll H_0$. 
%Initial and final states
%will be chosen to be eigenstates of $H_0$.
In a rearrangement reaction of the type $(A-X^-) + B\to C + X^-$,
where $C$ is a nuclear bound state between $A$ and $B$, the 
unperturbed and perturbed Hamiltonians for initial and final states
are different, i.e. $H_0^i\neq H_0^f$, $H_1^i\neq H_1^f$. In particular,
whereas in the initial state the perturbation is best chosen as the
nuclear attraction beween $A$ and $B$, i.e. $H_1^i = V_{NUC}$ and 
$H_0^i = H - H_1^i$, in the final state it will be the differential 
Coulomb force of $X^-$ on the nuclear bound state $C=(A-B)$. When
initial and final states are chosen as eigenstates to $H_0^i$ and $H_0^f$,
respectively, standard methods show that, in the Born approximation
the transition amplitude may be computed by either $\langle i|H_1^f f\rangle$
or $\langle f|H_0^i i\rangle$. The initial and final states are chosen as
\begin{eqnarray}
|i\rangle = |\Phi_{(A-X^-)}({\bf r}_A)\rangle\,\, |\Phi_{Coul}({\bf r}_B)\rangle \\
|f\rangle = |\Phi_{(A-B)}({\bf \rho})\rangle\,\, |\Phi_{Coul}({\bf s})\rangle
\end{eqnarray}
where ${\bf s}$ and ${\bf\rho}$ are the
$A-B$ center of mass and relative coordinates, respectively.
Coulomb wave functions $|\Phi_{Coul}\rangle$ and the
$A-X^-$ bound state wave function $\Phi_{(A-X^-)}$ were determined
numerically with realistic charge distributions. The nuclear wave
function $\Phi_{(A-B)}$ was parametrised by 
$\Phi = 2\sqrt{\gamma^5/3}\,\rho\,{\rm exp}(-\gamma\rho)$ with
$\gamma$ adjusted such that in the absence of $X^-$ the correct experimentally
determined cross section results.
The perturbation $H_1^f$ was chosen as the first non-vanishing element
in the expansion of the last two terms of Eq.~(\ref{Htot}) in terms of
relative coordinate $\rho$. For dipole transitions this results
into
\begin{equation}
H_1^f = -(Z_AR_A+Z_BR_B)e^2\frac{s_i\rho_i}{s^3}
\label{Hdipole}  
\end{equation}
whereas for quadrupole transitons
\begin{equation}
H_1^f = -(Z_AR_A^2+Z_BR_B^2)e^2\bigl(\frac{3}{2}\frac{s_is_j\rho_i\rho_j}{s^5}
- \frac{1}{2}\frac{\rho^2}{s^3}\bigr)
\label{Hquadro}
\end{equation}
where $R_A = M_B/(M_A+M_B)$ and $R_B = -M_A/(M_A+M_B)$. 
Rates were evaluated by numerical integration of the matrix elements
$\langle i|H_1^f f\rangle$ employing Fermi's Golden rule
\begin{equation}
\sigma v = \frac{2\pi}{\hbar}\, V \int{\rm d}N_f\,
\delta(E_i-E_f) |\langle i|H_1^f f\rangle |^2 
\end{equation} 
where $V$ is a normalization volume, $v$ relative velocity,
$\delta$ the Delta-function, and
\begin{equation}
{\rm d}N_f = \frac{V}{(2\pi\hbar)^3}p_C^2{\rm d}p_C{\rm d}\Omega_C
\end{equation}
a measure of the final phase space for nucleus C.
For the evaluation of the matrix elements, six-dimensional
integrals over the coordinates of two nuclei could be analytically
reduced to three-dimensional integrals which were numerically evaluated.
Similiar to Ref.~\cite{Hamaguchi} 
I have not considered internal spin of the nuclei, except
for the obvious total angular momentum degeneracy factors. 
Finally cross sections $\sigma (E)$ were converted to S-factors $S(E)$ .
They are related by
\begin{equation}
\sigma(E) = (S(E)/E)\,{\rm exp}(-G(E)), 
\end{equation}
where $E$ is center-of mass (CM) energy and ${\rm exp}(G)$ with 
\begin{equation}
G(E) = \frac{2\pi (Z_A-1)Z_B\alpha c}{v_{\rm CM}}
\label{GE}
\end{equation}
is the Coulomb repulsion factor. In the above $v_{\rm CM}$ 
is the relative velocity ($v_{\rm CM} \approx v_B$ for
bound states), and $\alpha$, $c$ fine structure constant and speed of light,
respectively. For assumptions concerning the angular momentum of the 
final $A$-$B$ nucleus, the number of multipoles included in the calculation,
and the assumed S-factor in the absence of bound states the reader is referred
to Table 2. 
The determined S-factors were
subsequently integrated over a thermal distribution to derive 
thermal nuclear rates
in the presence of bound states.


\begin{thebibliography}{99}
\bibitem{remark10} Reaction rates, for example,
for neutrons on \be7, are given by the product
of their reaction cross section $\sigma$ and the relative velocity 
between the reacting nuclei, $v$. It is
argued that due to the fast motion of the \be7
nucleus around the CHAMP, the relative velocity $v$ could be
enhanced by orders of magnitude. This is clearly not so, as
$v$ is, simply, part of the flux factor at large distances, defining the flux
of incoming neutrons in the cosmic rest frame, 
which is essentially
the rest frame of the heavy \be7-$X^-$ bound system. Due to the
small thermal velocities of the heavy bound state, the relative 
velocity, and thus the reaction rate, is even reduced, but only
about a factor of order unity. See also Ref.~\cite{Bird} 
for a more detailed
discussion.
\bibitem{remark20} The Bohr radius is only given for illustrative
purposes, all calculations assume bound state wave functions
determined by the Schroedinger equation, and deviating significantly
from the
$1s$ Bohr wave function. The approximative Bohr radius was determined
by requiring the same asymptotic slope at larger distances.
\bibitem{remark11} For example, the rapid \li6$(p,\alpha )$\he3 rate
for $T\simge 10\,$keV, continously destoys \li6-$X^-$ bound states,
such that the \li6 bound state fraction is below that expected
from naive estimates.
\bibitem{remark12} Throughout the paper, it is assumed that only
half of the stated $\Omega_X$ abundance is within negatively charged species
$X^-$, with the other half in the antiparticle of $X$, i.e. $X^+$,
such that electrical neutrality is kept.
\bibitem{remark13} Naively thought, some of the \li7 (synthesized as \be7) 
could survive
the rapid \be7(\h1$-X^- ,X^-){}^8$B destruction reaction at $T\approx 1\,$keV
if \be7 would be converted to \li7 at $\tau\simle 10^6$s. In fact, the
half life for \be7 against electron capture is 
$\tau_{\rm {}^7Be}\approx 4.6\times 10^6$s. Nevertheless, the binding
energy of an electon to the \be7 nucleus is $E_b\approx 0.21\,$keV, such that
the \be7 nucleus stays bare until much lower temperatures $T\ll 1\,$keV.
Since the electron plasma density is well below the effective density of
electrons in a recombined \be7 nucleus, the \be7 nuclei has to either avait
recombination or $\beta^+$-decay to convert to \li7. Due to the rapidness
of the \be7 destruction process, it is therefore expected that \be7 $\to$\li7
at low temperatures plays hardly a role.
\bibitem{remark301}
The effects of annihilation of $X^-$ with $X^+$ in $X^--X^+$ bound states 
forming to some degree~\cite{Bird} 
are subdominant compared to those of the decay
itself.
\end{thebibliography}
\end{document}